\documentclass[twocolumn,tradiabstract]{aa}
\usepackage{graphicx}
\usepackage{amsbsy}
\usepackage{natbib}
\usepackage{color}
\usepackage{txfonts}
\usepackage{url}

\newcommand{\vc}[1]{{\boldsymbol #1}}
\newcommand{\de}{\mathrm{d}}

\DeclareMathSymbol{\varOmega}{\mathord}{letters}{"0A}
\DeclareMathSymbol{\varSigma}{\mathord}{letters}{"06}
\DeclareMathSymbol{\varPsi}{\mathord}{letters}{"09}

\newcommand{\Eq}[1]{Eq.~(\ref{#1})}

\newcommand{\App}[1]{Appendix~\ref{#1}}
\newcommand{\Sec}[1]{Sect.~\ref{#1}}

\newcommand{\Fig}[1]{Fig.~\ref{#1}}

\begin{document}

\title{Adding particle collisions to the formation of\\asteroids and Kuiper
belt objects via streaming instabilities}
\titlerunning{Particle collisions and the formation of asteroids and Kuiper
belt objects}

\author{Anders Johansen\inst{1}, Andrew N.\ Youdin\inst{2}, \and Yoram
Lithwick\inst{3}}
\authorrunning{Johansen, Youdin, \& Lithwick} 

\offprints{\\A.\ Johansen (\email{anders@astro.lu.se})}

\institute{Lund Observatory, Department of Astronomy and Theoretical Physics,
Lund University, Box 43, 221 00 Lund, Sweden \and Harvard-Smithsonian Center
for Astrophysics, 60 Garden Street, Cambridge, MA 02138, USA \and Center for
Interdisciplinary Exploration and Research in Astrophysics (CIERA) and
Dept.\ of Physics \& Astronomy, Northwestern University, 2145 Sheridan
Rd, Evanston, IL 60208, USA}

\abstract{Modelling the formation of super-km-sized planetesimals by
gravitational collapse of regions overdense in small particles requires
numerical algorithms capable of handling simultaneously hydrodynamics, particle
dynamics and particle collisions. While the initial phases of radial
contraction are dictated by drag forces and gravity, particle collisions become
gradually more significant as filaments contract beyond Roche density. Here we
present a new numerical algorithm for treating momentum and energy exchange in
collisions between numerical superparticles representing a high number of
physical particles. We adopt a Monte Carlo approach where superparticle pairs
in a grid cell collide statistically on the physical collision time-scale.
Collisions occur by enlarging particles until they touch and solving for the
collision outcome, accounting for energy dissipation in inelastic collisions.
We demonstrate that superparticle collisions can be consistently implemented at
a modest computational cost. In protoplanetary disc turbulence driven by the
streaming instability, we argue that the relative Keplerian shear velocity
should be subtracted during the collision calculation. If it is not
subtracted, density inhomogeneities are too rapidly diffused away, as bloated
particles exaggerate collision speeds. Local particle densities reach several
thousand times the mid-plane gas density. We find efficient formation of
gravitationally bound clumps, with a range of masses corresponding to
contracted radii from 100 to 400 km when applied to the asteroid belt and 150
to 730 km when applied to the Kuiper belt, extrapolated using a constant
self-gravity parameter. The smaller planetesimals are not observed at low
resolution, but the masses of the largest planetesimals are relatively
independent of resolution and treatment of collisions.}


\maketitle

\section{Introduction}

The formation of super-km-sized planetesimals is an important step towards
terrestrial planets and the solid cores of gas and ice giants
\citep[e.g.][]{Safronov1969,Goldreich+etal2004,ChiangYoudin2010}. The asteroid
and Kuiper belts of the solar system, as well as the extrasolar debris discs,
are believed to be left-over populations of planetesimals that did not grow to
planets. Comparing models and simulations of planetesimal formation to
observations of such planetesimal belts constrains our theoretical picture of
the planetesimal formation stage, and at the same time it gives insight into
the physical processes that shaped the architectures of these systems
\citep{Morbidelli+etal2009,Weidenschilling2010,Nesvorny+etal2010,SheppardTrujillo2010,Krivov2010,KenyonBromley2010}.

Planetesimal formation takes place in a complex environment of turbulent gas
interacting via drag forces with particles of many sizes. The streaming
instability thrives in the systematic relative motion of gas and particles and
leads to spontaneous clumping of particles
\citep{YoudinGoodman2005,JohansenYoudin2007,BaiStone2010b}, seeding a
gravitational collapse into bound clumps \citep{Johansen+etal2009} and further
to solid planetesimals \citep{Nesvorny+etal2010}. While the latest years have
seen major progress in numerical modelling of drag force interaction between
particles and gas
\citep{YoudinJohansen2007,Balsara+etal2009,Miniati2010,BaiStone2010a} as well
as the self-gravity of the particle layer
\citep{Johansen+etal2007,Rein+etal2010}, good algorithms for treating
simultaneously hydrodynamics, gravitational dynamics and particle collisions
are still missing.

There are two main approaches in astrophysics to treating particle collisions
in numerical simulations. Modelling a set of {\it physical particles} with
collision tracking allows simulation of particle aggregation in close
concordance with the nature of real physical collisions. This method has
successfully been applied to model the particle rings of Saturn
\citep{WisdomTremaine1988,Salo1991,KarjalainenSalo2004} and to model collisions
between individual dust grains and aggregates \citep{DominikNubold2002}. The
drawback of the physical-particle approach is that the size of the system is
limited by the number of numerical particles that can be afforded in the
simulation. The formation of a Ceres-mass planetesimal from 10-cm-sized rocks
would e.g.\ require tracking of $\mathcal{O}(10^{20})$ particles, orders of
magnitude beyond what current computational resources allow.

Algorithms involving {\it inflated particles} group collections of physical
particles into much larger numerical particles under conservation of total mass
$M$ and mean free path $\lambda$. Decreasing the particle number $N$ to a
number that can be handled in a computer simulation, while maintaining
$\lambda^{-1} \equiv (N/V) \sigma$ by artificially increasing the collisional
cross section $\sigma$, yields the correct collision frequency in systems that
are much larger than what can be resolved with the physical particle approach.
The inflated particle approach was used recently by \cite{LithwickChiang2007},
\cite{Michikoshi+etal2007}, \cite{Nesvorny+etal2010}, and \cite{Rein+etal2010},
with different methods for tracking the actual collision, but the concept of
bloated particles has deeper roots \citep[e.g.][]{KokuboIda1996}.

In this paper we put forward a new algorithm to model collisions between
numerical {\it superparticles}. Superparticles are designed to represent swarms
of physical particles. The aerodynamical properties of the superparticle (e.g.\
the friction time) is still that of a single physical particle. Superparticles
are widely used to model the solid particle component in computer simulations
of coupled gas and particle motion in protoplanetary discs
\citep{JohansenYoudin2007,BaiStone2010b}. Since superparticles can be
considered to represent swarms of smaller particles, direct collision tracking
is not possible. \cite{Johansen+etal2007} modelled superparticle collisions by
damping the random motion of particles inside a grid cell on the collisional
time-scale. They showed that inelastic collisions, where part of the kinetic
energy is converted to heat and deformation during the collisions, is
beneficial for the gravitational collapse and allows the formation of
planetesimals in protoplanetary discs of lower mass, compared to simulations
without damping. However, the simplified collision scheme of
\cite{Johansen+etal2007} is insufficient in capturing the pairwise momentum
exchange and energy dissipation.

We develop here a statistical approach to model the full momentum exchange and
energy dissipation in collisions between superparticles. The Monte Carlo scheme
is inspired by the collision algorithms presented by \cite{LithwickChiang2007}
and \cite{ZsomDullemond2008}. The essence of our algorithm is to determine the
collision time-scale between all superparticle pairs within a grid cell. Two
superparticles collide as if they were physical particles touching each other,
if a random number chosen uniformly between zero and one is smaller than the
ratio of the simulation time-step to the collision time-scale.

Collisions can be followed together with hydrodynamics at a moderate
computational cost depending only on the number of particles per grid cell. We
compare the statistical properties of the particle density in 3-D
hydrodynamical simulations with and without collisions. Including the
self-gravity of the particles, we find formation of bound clumps, with masses
comparable to that of the 500-km-radius dwarf planet Ceres when applied to the
asteroid belt, relatively independently of numerical resolution and treatment
of collisions. The scale-free nature of our simulations allows application of
the results to the Kuiper belt as well, with contracted planetesimal radii
approximately 80\% higher than in the asteroid belt.

The paper is organised as follows. In \Sec{s:algorithm} we describe the new
superparticle collision algorithm. The algorithm is tested against known test
problems and conservation properties of the shearing box in
\Sec{s:validation}. In \Sec{s:streaming} we analyse statistical properties of
the particle density achieved in simulations of gas and particle turbulence
driven by the streaming instability. We continue to include self-gravity in the
simulations and analyse the planetesimal masses obtained under various
assumptions about collisions in \Sec{s:planetesimals}. We summarise and discuss
our results in \Sec{s:discussion}. The appendices A--C contain further
descriptions of the collision algorithm.

\section{Superparticle collision algorithm}
\label{s:algorithm}

We will use the notation that a superparticle represents a swarm of physical
particles with number density $\hat{n}$ and volume $\delta V$. Since we are
interested in coupling superparticle collisions to grid hydrodynamics, the
volume is taken to be that of a grid cell, $\delta V = \delta x \times \delta y
\times  \delta z$. The physical particles in the swarm have individual mass,
physical radius, material density, and collisional cross section $m$, $R$,
$\rho_\bullet$ and $\sigma$. We assume that all swarms are similar, both in
internal particle number and in the physical mass of the constituent particles.

To track a collision we calculate the mean free path $\hat{\lambda}$ for a test
particle interacting with the swarm of particles represented by a single
superparticle,
\begin{equation}
  \hat{\lambda} = \frac{1}{\hat{n}\sigma} \, .
\end{equation}
Superparticles in the same grid cell are considered as potential colliders. For
each collision pair the collision time-scale is calculated from
\begin{equation}
  \tau_{\rm c} = \frac{\hat{\lambda}}{\delta v} \, ,
\end{equation}
where $\delta v$ is the relative speed between particles $i$ and $j$. The
simulation time-step $\delta t$, set by hydrodynamics and drag forces, is then
used to calculate the probability that those two particles collide in this
time-step,
\begin{equation}
  P = \frac{\delta t}{\tau_{\rm c}} \, .
  \label{eq:pcoll}
\end{equation}

Two colliding swarms have their velocity vectors changed instantaneously. The
collision outcome is found by considering two virtual spherical particles whose
surfaces touch, with particle centres at the locations of the superparticles,
and solving for momentum conservation and inelastic energy dissipation (or
energy conservation, in case of elastic collisions). We define the velocity
vectors relative to the mean velocity field
$\overline{\vc{v}}=(\vc{v}_j+\vc{v}_k)/2$,
\begin{eqnarray}
  \vc{v}'_j &=& \vc{v}_j - \overline{\vc{v}} \, , \\
  \vc{v}'_k &=& \vc{v}_k - \overline{\vc{v}} = -\vc{v}'_j \, . 
\end{eqnarray}
Here $\vc{v}_j$ and $\vc{v}_k$ are the velocity vectors of the two
particles\footnote{We show in \Sec{s:shear} that the Keplerian shear should be
subtracted from the velocity vectors when determining both the collision
time-scale and the collision outcome, in the limit of particles that are much
smaller than a grid cell.}. The normal vector $\vc{e}_\perp$ connecting the
centres of the particles at the time of collision is calculated as
\begin{equation}
  \vc{e}_\perp = \frac{\vc{x}_j-\vc{x}_k}{|\vc{x}_j-\vc{x}_k|} \, .
\end{equation}
The parallel vector $\vc{e}_\parallel$ is perpendicular to $\vc{e}_\perp$ in
the same plane as the relative velocity vector. The relative velocity vectors
are now decomposed on the two directions
\begin{eqnarray}
  \vc{v}'_j &=& a_j \vc{e}_\perp + b_j \vc{e}_\parallel \, , \\
  \vc{v}'_k &=& a_k \vc{e}_\perp + b_k \vc{e}_\parallel \, ,
\end{eqnarray}
with $a_k=-a_j$ and $b_k=-b_j$. In the collision we maintain $b$, while we
reflect $a$ according to
\begin{equation}
  a \rightarrow -\epsilon a \, .
\end{equation}
Here $\epsilon\in[0,1]$ is the coefficient of restitution, parameterising the
degree of energy dissipation during the collision. Inelastic collisions can
play an important role in dissipating kinetic energy and facilitating the
gravitational collapse phase. In general the coefficient of restitution depends
on material parameters, impact speed and ambient temperature. Water ice
particles have been measured to have a high coefficient of restitution
$\epsilon\approx0.9$ for impact speeds below $\approx 2$ m/s
\citep[quasi-elastic regime of][]{Higa+etal1996}. Above this critical speed the
measured coefficient of restitution rapidly drops towards zero. More recent
microgravity and drop tower experiments find a coefficient of restitution
between 0.06 and 0.84 in low-velocity collisions between 1.5-cm-sized icy
pebbles \citep{Heisselmann+etal2010}. In this paper we consider for the sake of
simplicity the coefficient of restitution to be a constant that is independent
of the relative speed.

The collision time-scale has a simple relation to the friction time-scale when
particles are small and drag forces are in the Epstein regime. We show in
\App{s:taucfromtauf} how the collision time-scale can be easily calculated from
the friction time-scale, useful e.g.\ for simulations of gas and particles in
protoplanetary discs.

Consider now a grid cell containing $N$ superparticles. For particle $i$ the
collision probability for a representative
particle\footnote{\cite{ZsomDullemond2008} define a representative particle
from a swarm as a test particle (a random particle from the swarm) used to
probe the collision time-scale with another swarm.} from superparticle $i$ to
collide with the particle swarms $j=i+1$ to $j=N$ is calculated. The collision
occurs if a random number, drawn for each collision partner, is smaller than
$P$ from \Eq{eq:pcoll}. The collision instantaneously changes the velocity
vectors of both particles $i$ and $j$. This way the correct collision frequency
is obtained for both particles, even though the algorithm only considers the
possible collision $i$ with $j$, but not $j$ with $i$. In \App{s:limiter} we
describe how to consistently limit the number of collision partners, and thus
save computation time, in grid cells which contain many ($\gg$ 100) particles.

There are several advantages to using such a probabilistic swarm approach to
particle collisions. We mention here a few: (i) it is fast because we do not
have to track when particles touch or overlap within the grid cells, (ii) it
allows us to freely choose the relative speed that enters the collision
frequency, useful e.g.\ for subtracting off the Keplerian shear (see Sect.\
\ref{s:shear}), and (iii) the algorithm is easily generalisable to also include
a probabilistic approach to particle coagulation and shattering.

\begin{figure}
  \includegraphics[width=8.7cm]{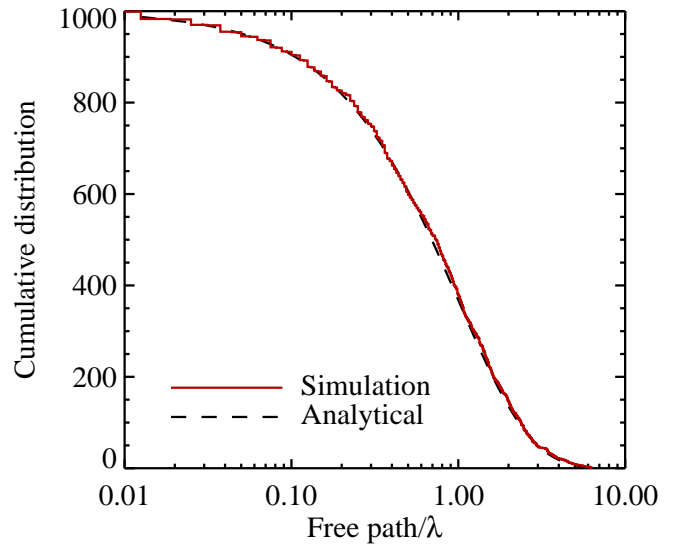}
  \caption{Cumulative free path for 1000 superparticles released into medium
  with mean-free-path of $\lambda=0.1$. The distribution function follows the
  analytical expectation $N=N_0 \exp(-\ell/\lambda)$ very closely. Our Monte
  Carlo algorithm for superparticle collisions gives a free path in good
  agreement with the real physical system consisting of many more particles.}
  \label{f:freepath_dist}
\end{figure}
In \Fig{f:freepath_dist} we show the collision path length of test particles
injected into a medium with 10 superparticles per grid cell and a mean free
path of $\lambda=0.1$. Collisions are tracked through the Monte Carlo method
described above. The collision algorithm makes some particles collide after a
short flight path and others after a longer. The distribution plotted in
\Fig{f:freepath_dist} follows closely the expectation $N=N_0
\exp(-\ell/\lambda)$. The Monte Carlo approach to collisions is very similar to
the physical particle approach in the distribution of free flight paths.

The main technical difference between using inflated particles (see
introduction) and our newly developed collision algorithm for superparticles is
that inflated particles always collide when they overlap physically (the
particle size can be associated with the grid cell size), while superparticles
sharing the same grid cell collide with a certain probability which guarantees
that collisions occur on the average after a collisional time-scale. Another
difference is that superparticles which do not approach must still be allowed
to collide, as otherwise the mean free path will be too long. Non-approaching
particles are collided by flipping the relative velocity vector before
collision and reflipping afterwards. The main issue with approaching collisions
is that collisions occur in fixed grid cells which are not centred on the
superparticle in question, and thus a superparticle at the edge of a grid cell
will have too few collision partners if only approaching collisions are
allowed. We show in \App{s:inflated} how the superparticle approach transforms
smoothly to the inflated particle approach when the number of superparticles is
reduced.

The Monte Carlo collision scheme presented here could equally well be
formulated in terms of inflated particles, by constructing inflated particles
smaller than a grid cell. Solving statistically for the collision outcome of
these ``sub-grid'' particles is mathematically equivalent to the
interpretation, chosen for this paper, of the numerical particles as swarms.

\section{Validation of algorithm}
\label{s:validation}

We have implemented the Monte Carlo superparticle collision scheme described in
\Sec{s:algorithm} into the open source code Pencil Code\footnote{The
code, including the developments described in this paper, can be freely
downloaded at \url{http://code.google.com/p/pencil-code/}.}. The Pencil Code
evolves gas on a fixed grid and has fully parallelised modules for an
additional solid component represented by superparticles
\citep{Johansen+etal2007,YoudinJohansen2007}. We first validate the collision
algorithm in the limit of inflated particles (i.e.\ where two particles
occupying the same grid cell always collide and only approaching collisions are
considered), to compare our results directly to those of
\cite{LithwickChiang2007}. The 2-D algorithm of \cite{LithwickChiang2007} has a
probabilistic approach to determine whether two particles are in the same
vertical zone when they overlap in the plane. Their algorithm can thus be seen
as a hybrid of the inflated particle approach and a Monte Carlo scheme.

We set up a test problem similar to the one presented in
\cite{LithwickChiang2007}. We define a 2-D simulation box covering the spatial
interval $[-2,+2]\times[-2,+2]$ with $4000$ grid cells in both the $x$ and $y$
direction. $10^4$ particles are placed randomly in a ring of full width $0.08$
centred at the radial distance $r=1$. A central gravity source, of strength $G
M=1$, is placed in the centre of the coordinate frame.

We integrate the particle orbits, including collisions, for $10^4$ revolutions
of the ring centre. In order to compare directly with \cite{LithwickChiang2007}
we use their 2-D approximation. The particle number density can be approximated
as $n \sim \varSigma/H$, where $\varSigma$ is the column (number) density and
$H$ is the scale height of the particle disc. The random particle motion $u$
can be written as $u \sim H \varOmega$. This yields a collision time
\begin{equation}
  \tau_{\rm c}^{{\rm (2D)}} \sim \frac{1}{n \sigma u} \sim \frac{1}{\varSigma
  \sigma \varOmega} \sim \frac{T_{\rm orb}}{\tau} \, ,
\end{equation}
where $\tau=\varSigma \sigma$ is the vertical optical depth of the disc and
$T_{\rm orb}=2\pi/\varOmega$ is the orbital time-scale. While the collision
time-scale in general depends on the random particle motion, this dependence
vanishes in the 2-D Keplerian disc approximation -- faster random motion
cancels with increased particle scale-height in the collision time expression.

Requiring that orbits are maintained for $10^4$ orbital time-scales, we set the
time-step of the Pencil Code to $\delta t=0.01 \varOmega^{-1}$, covering each
orbit $T_{\rm orb}=2\pi/\varOmega$ by around 600 time-steps. This proved
necessary because the third order time integration scheme of the Pencil Code is
not constructed to conserve orbital angular momentum and energy. Using the
highly optimized orbital dynamics code SWIFT, \cite{LithwickChiang2007} solve
the same problem with slightly less than five time-steps per orbit.

\begin{figure}
  \includegraphics[width=8.0cm]{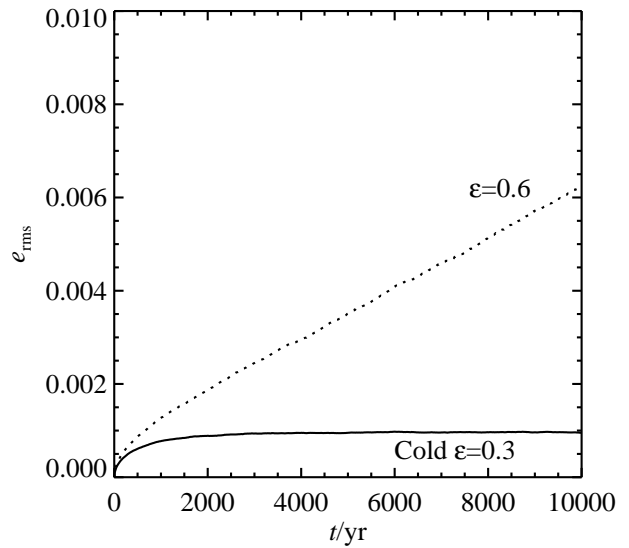}
  \caption{The eccentricity evolution of particles orbiting a central gravity
  with $G M=1$. Relatively inelastic collisions, with coefficient of
  restitution $\epsilon=0.3$, evolve towards an equilibrium eccentricity of
  $10^{-3}$, with orbital excursions comparable to the grid spacing. More
  elastic collisions, with $\epsilon=0.6$, lead to catastrophic heating of the
  particle system. The results follow closely Fig.\ 1 of
  \cite{LithwickChiang2007}.}
  \label{f:eccrms_t}
\end{figure}

In \Fig{f:eccrms_t} we show the eccentricity evolution of the particle ring.
For a coefficient of restitution of $\epsilon=0.3$ the particles relax to an
equilibrium eccentricity of around $e_{\rm rms}=0.001$, comparable to $\delta
x/r$. A higher coefficient of restitution of $\epsilon=0.6$ leads instead to
catastrophic heating of the disc \citep{GoldreichTremaine1978}, with an
eccentricity that evolves linearly with time. The results presented in
\Fig{f:eccrms_t} show that the superparticle collision algorithm is in
excellent agreement with \cite{LithwickChiang2007} in the limit of inflated
particles.

\subsection{Density evolution}

The width of a particle ring increases due to collisional viscosity. Since the
collision time-scale scales inversely with particle density, the collisional
evolution slows down with time. An analytical solution to the diffusion problem
was found by \cite{PetitHenon1987}. In the notation of
\cite{LithwickChiang2007} the width $\sigma_r$ of an initially narrow ring
increases according to
\begin{equation}
  \sigma_r = \left( \frac{36}{20^{3/2}} k_\nu \frac{(\delta x)^4}{\overline{r}}
  N_{\rm tp} \frac{t}{T_{\rm orb}} \right)^{1/3} \, .
\end{equation}
Here $k_\nu$ is a dimensionless factor that depends on the coefficient of
restitution $\epsilon$, $\delta x$ is the grid spacing, $\overline{r}$ is the
mean radial coordinate of the particles, $N_{\rm tp}$ is the particle number
and $t$ the time.

\begin{figure}
  \includegraphics[width=8.0cm]{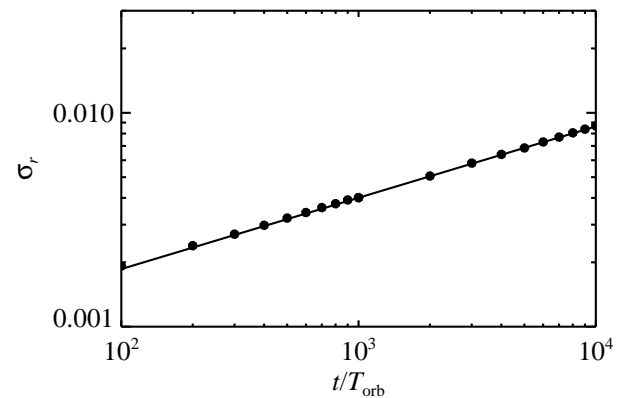}
  \caption{The width of a particle ring orbiting a central gravitating mass
  versus time. The 10000 particles were initially placed in a ring centred at
  $r=1$ and a width of $2 \Delta = 10^{-3}$, similar to the grid spacing.
  Compare to upper panel of Fig.\ 3 in \cite{LithwickChiang2007}.}
  \label{f:rprms_t}
\end{figure}

We follow \cite{LithwickChiang2007} and define an initially very narrow ring of
radial extent $2 \Delta=10^{-3}$. The units follow from our choice of $G M=1$.
The evolution of the radial width is shown in \Fig{f:rprms_t} over $10^4$
orbits. We overplot the analytical solution for $k_\nu=0.016$, similar to the
fit in \cite{LithwickChiang2007}, and find excellent agreement.

\subsection{Superparticle collisions in the local frame}

Hill's equations describe motion relative to a frame that corotates with the
Keplerian frequency $\varOmega$ at an arbitrary distance from the central
gravity source. The coordinate axes are defined such that $x$ points radially
outwards and $y$ points along the flow of the disc. The 2-D equations of motion
of particles are
\begin{eqnarray}
  \frac{\de v_x}{\de t} &=& +2 \varOmega v_y + 3 \varOmega^2 x \, , \\
  \frac{\de v_y}{\de t} &=& -2 \varOmega v_x \, .
\end{eqnarray}
Particle positions are evolved through $\dot{\vc{x}}=\vc{v}$. The boundary
conditions are periodic in the azimuthal direction. Particles passing over the
inner (outer) radial boundary get the velocity $(3/2)\varOmega L_x$ subtracted
(added) to their azimuthal velocity. We also refer to the frame as the shearing
box. We consider a box size of $L_x=L_y=0.2$ covered by $32^2$ grid cells and
$102{,}400$ particles.

\begin{figure}
  \includegraphics[width=8.0cm]{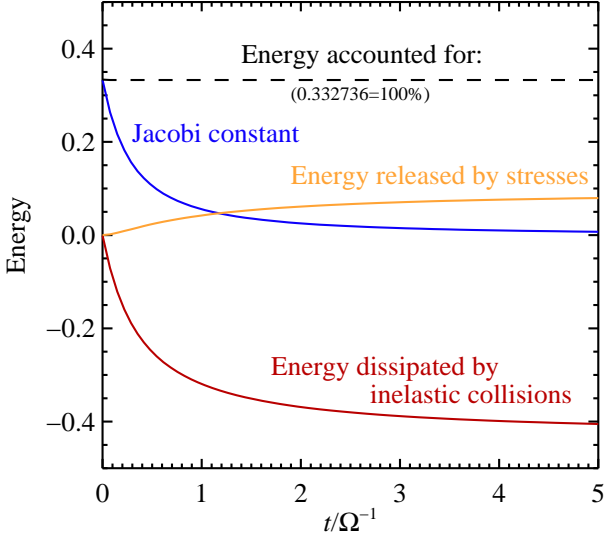}
  \caption{Evolution of energy in a shearing box simulation where particles
  have a mean-free-path of $\lambda=0.1 H$ and coefficient of restitution
  $\epsilon=0.3$. Drag forces are ignored. The Jacobi constant falls due to
  dissipative collisions. By monitoring the energy released as particles pass
  the boundaries and the energy dissipation by inelastic collisions we can
  account for all the energy in the system.}
  \label{f:energy_t}
\end{figure}

The conserved energy (Jacobi constant) is
\begin{equation}
  E = \frac{1}{2} m \dot{x}^2 + \frac{1}{2} m \dot{y}^2 - \frac{3}{2} m
  \varOmega^2 x^2 \, .
\end{equation}
Elastic collisions re-orient the particles without changing energy, and thus
convert circular orbits into eccentric ones while conserving energy. Ignoring
gas, which damps the velocity relative to the gas and hence the eccentricity,
elastic collisions conserve the Jacobi energy. \Fig{f:energy_t} shows the
energy of particles versus time in local frame simulation with inelastic
collisions. Particles are initialised with random position and velocity vectors
($\delta v=1$). The mean-free-path is $\lambda=0.1 H$, giving an initial
collision time-scale of $\tau_{\rm c} \sim 0.1$. The coefficient of restitution
is $\epsilon=0.3$. The Jacobi constant falls with time due to the energy
dissipated by inelastic collisions. At the same time particles passing over the
radial boundaries release energy from the Keplerian shear through their mean
Reynolds stress (the code tracks and outputs that energy release for each
particle passing the radial boundary). All energy in the system is accounted
for in these three reservoirs.
\begin{figure}
  \includegraphics[width=8.0cm]{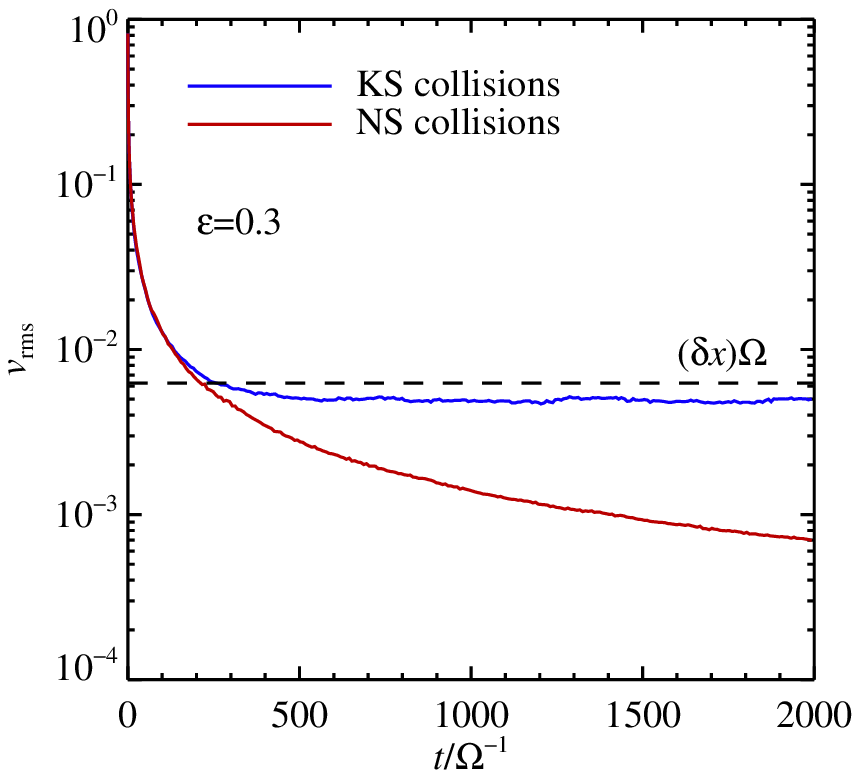}\\
  \includegraphics[width=8.0cm]{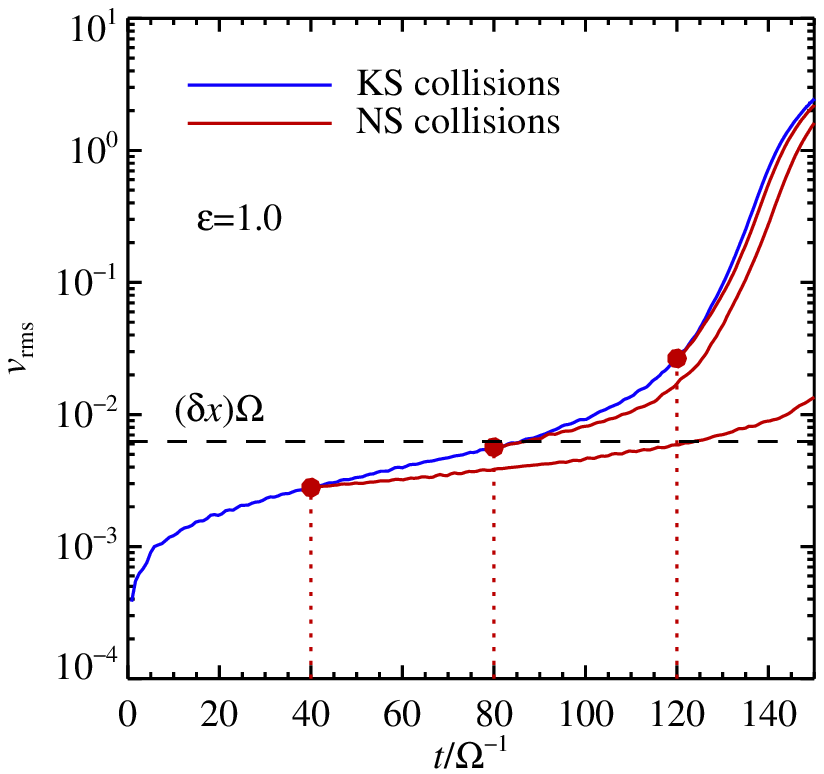}
  \caption{Evolution of particle rms speed in the shearing box for a simulation
  with normal collisions (KS, blue/black line) and a simulation in which the
  relative Keplerian shear is subtracted when determining the collision
  time-scale and outcome (NS, red/gray line). The top panel shows the decay of
  initially random particle motion due to inelastic collisions
  ($\epsilon=0.3$). The rms speed can not fall below $v_{\rm rms}\approx(\delta
  x)\varOmega$ for KS collisions, due to the energy release from the Keplerian
  shear. In the simulation with NS collisions, on the other hand, the rms speed
  continues to decay towards zero. In the bottom panel we consider elastic
  collisions ($\epsilon=1.0$) with zero random motion initially. Energy is
  released from the Keplerian shear. The blue line shows results of simulations
  with NS collisions, rerun from snapshots of the KS simulation at various
  times. The two solutions match increasingly well when the particle rms speed
  increases above $(\delta x)\varOmega$.}
  \label{f:vprms_t_shear_comparison}
\end{figure}

\subsubsection{Shear during collision}
\label{s:shear}

Particle collisions in the shearing box release energy from the Keplerian shear
into random motion, leading in the absence of drag forces either to
catastrophic heating ($v_{\rm rms} \rightarrow \infty$) or to an equilibrium
with energy dissipation in inelastic collisions ($v_{\rm rms}$$\sim$$R
\varOmega$ where $R$ is the particle radius). Discounting the former option,
the result of the latter can be artificially exaggerated by the numerical
scheme because we identify the collision between two superparticle swarms with
the collision between two members of the swarms located at the respective swarm
centres. In reality collisions would occur between neighbouring particles
separated by less than their physical diameter. The naive numerical algorithm
would make the system settle for an equilibrium where $v_{\rm rms}$$\sim
$$(\delta x) \varOmega$, where $\delta x$ is the grid spacing and also the
typical distance between superparticle centres. This rms speed greatly exceeds
the desired $v_{\rm rms}\sim R \varOmega$. In other words, the naive collision
algorithm will input artificial heating.
\begin{table*}
  \caption{Simulation parameters.}
  \begin{center}
    \begin{tabular}{lcccccccc}
      \hline
      \hline
      Run & $L_x \times L_y \times L_z$
          & $N_x \times N_y \times N_z$
          & $N_{\rm par}$
          & $\varOmega\tau_{\rm f}$
          & Collisions
          & $\epsilon$
          & $\Delta t$
          & $t_{\rm sg}$ \\
      \hline
      SI64\_nocoll    & $0.2\times0.2\times0.2$ & $64\times64\times64$
                      &  $300{,}000$    & $0.3$ & -- & --    & 100 & --  \\
      SI64\_e1.0      & $0.2\times0.2\times0.2$ & $64\times64\times64$
                      &  $300{,}000$    & $0.3$ & KS & $1.0$ & 100 & --  \\
      SI64\_e0.3      & $0.2\times0.2\times0.2$ & $64\times64\times64$
                      &  $300{,}000$    & $0.3$ & KS & $0.3$ & 100 & --  \\
      SI64\_e0.3\_NS  & $0.2\times0.2\times0.2$ & $64\times64\times64$
                      &  $300{,}000$    & $0.3$ & NS & $0.3$ & 100 & 52 \\
      SI128\_nocoll   & $0.2\times0.2\times0.2$ & $128\times128\times128$
                      & $2{,}400{,}000$ & $0.3$ & -- & --    & 50 & --  \\
      SI128\_e1.0     & $0.2\times0.2\times0.2$ & $128\times128\times128$
                      & $2{,}400{,}000$ & $0.3$ & KS & $1.0$ & 50 & --  \\
      SI128\_e0.3     & $0.2\times0.2\times0.2$ & $128\times128\times128$
                      & $2{,}400{,}000$ & $0.3$ & KS & $0.3$ & 50 & --  \\
      SI128\_e0.3\_NS & $0.2\times0.2\times0.2$ & $128\times128\times128$
                      & $2{,}400{,}000$ & $0.3$ & NS & $0.3$ & 50 & 19 \\
    \end{tabular}
  \end{center}
  Col.\ (1): Name of simulation. Col.\ (2): Box size in scale heights. Col.\
  (3): Resolution. Col.\ (4): Number of particles. Col.\ (5): Friction time.
  Col.\ (6): Collision type. Col.\ (7): Coefficient of restitution. Col.\ (8):
  Simulation time in orbits. Col.\ (9): Time of starting self-gravity.
  \label{t:parameters}
\end{table*}

Collisions between particles of radius $R \ll \delta x$ can be modelled by
subtracting the Keplerian shear part from the relative speed both for
determining the collision time-scale and for determining the outcome of the
collision. Decomposing the azimuthal velocity field as $\dot{y}=\tilde{v}_y +
v_y^{(0)}$, where $v_y^{(0)}=-(3/2) \varOmega x$ is the Keplerian shear
velocity and $\tilde{v}_y$ is the peculiar velocity, we can calculate both the
collision time-scale and outcome in terms of $\tilde{v}_y$ (together with $v_x$
and $v_z$). \cite{Lyra+etal2009} applied a similar trick to subtract off the
entire (Keplerian plus peculiar) gas velocity from the particle velocity.
However, two particles moving at the same velocity as the local gas do not
necessarily avoid collisions, even if the gas is incompressible, since the
particle motion is not completely coupled to the gas. Therefore we choose in
this paper to subtract off only the Keplerian orbital speed from the particle
velocity. The dynamical equations of the Pencil Code are already formulated
relative to the Keplerian shear, so subtracting off the shear is natural to the
governing system of equations.

Collisions relative to the Keplerian shear conserve both the total momentum and
the momentum relative to the Keplerian shear, but the energy in elastic
collisions is only conserved relative to the Keplerian shear. To see this,
consider the kinetic energy of two particles,
\begin{equation}
  E =
  \frac{1}{2} m \left\{ v_{x1}^2 + [\tilde{v}_{y1}+v_{y1}^{(0)}]^2 \right\} + 
  \frac{1}{2} m \left\{ v_{x2}^2 + [\tilde{v}_{y2}+v_{y2}^{(0)}]^2 \right\} \, .
\end{equation}
Here $m$ is the mass of a superparticle, assumed to be the same for both
colliders. An elastic collision solved in terms of ($v_{x1}$, $\tilde{v}_{y1}$,
$v_{x2}$, $\tilde{v}_{y2}$) conserves both the sum of the squares of those
velocity components, as well as the squares of $v_{y1}^{(0)}$ and
$v_{y2}^{(0)}$ (the latter is true since the position $x$ is not changed by the
collision). The difference in energy before and after the collision is therefore
\begin{equation}
  \Delta E = E_{\rm after} - E_{\rm before} = m [
  \Delta \tilde{v}_{y1} v_{y1}^{(0)} + \Delta \tilde{v}_{y2} v_{y2}^{(0)} ] \, .
\end{equation}
This result holds also in 3-D. The energy difference is generally not zero,
even though $\Delta \tilde{v}_{y1}=-\Delta \tilde{v}_{y2}$ by momentum
conservation, since the offset $v_y^{(0)}$ is not the same for the two
particles. The non-conservation is nevertheless small: the azimuthal velocity
change in the collision is uncorrelated with the Keplerian shear velocity, so
$\langle \Delta \tilde{v}_y v_y^{(0)} \rangle_{\rm box}\approx0$. The particle
integrator's slight non-conservation of Keplerian orbits is not a serious
limitation in simulations where the dynamics is driven by hydrodynamical
instabilities and drag forces. The correct relative Keplerian shear based on
the physical size of the particles can in principle be added artificially, to
obtain the correct energy release from the shear, but this is negligible for
1--10 cm particles considered in this paper.

The total angular momentum of two colliding particles,
\begin{equation}
  \vc{L} = m \vc{r}_1 \times \vc{v}_1 + m \vc{r}_2 \times \vc{v}_2 \, ,
\end{equation}
is conserved in the collisions, both with and without Keplerian shear in the
collision, as long as the force during the collision acts along the line
connecting the two particles. This is the case both with and without Keplerian
shear. For equal-mass particles we can write the change in the velocity as
$\Delta \vc{v}_1 = -\Delta \vc{v}_2 = c (\vc{r}_2-\vc{r}_1)$, giving
\begin{equation}
  \Delta \vc{L} = m \vc{r}_1 \times \Delta \vc{v}_1 + m \vc{r}_2 \times
  \Delta \vc{v}_2 = 0 \, .
\end{equation}
The above arguments for energy and angular momentum conservation are
generalisible to distinct particle masses as well. However, while the Monte
Carlo collision scheme in itself is fully consistent with distinct particle
masses, correct energy equipartition among particle sizes can not be obtained
with equal-mass superparticles (see discussion in \App{s:multiple}).

In the following we use the abbreviations KS for collisions that include
Keplerian shear and NS for collisions where the Keplerian shear is subtracted
off when determining the collision time-scale and outcome.
\Fig{f:vprms_t_shear_comparison} shows the evolution of the particle rms speed
in a shearing box simulation. The top panel shows the decay of initially random
particle motion by inelastic ($\epsilon=0.3$) collisions for KS collisions and
for NS collisions. KS collisions decay towards $v_{\rm rms}\approx(\delta
x)\varOmega$, the random motion released by the Keplerian shear in a single
collision. NS collisions on the other hand continue to decay towards zero. In
the bottom panel of \Fig{f:vprms_t_shear_comparison} we start with zero random
motion and observe how elastic ($\epsilon=1.0$) KS collisions heat up the
system. Rerunning the simulation with elastic NS collisions from various
starting times of the KS simulation shows clearly that the evolution of the
system is very similar as long as the particle rms speed is larger than
$(\delta x) \varOmega$. In actual simulations with gas and hydrodynamical
instabilities driving particle dynamics with characteristic motion much faster
than $v\sim(\delta x)\varOmega$, one can subtract off the Keplerian shear term
when determining the time-scale and outcome of collisions and still model the
correct system, without any spurious energy released by bloated particles.

\section{Particle collisions and the streaming instability}
\label{s:streaming}

Armed with a collision algorithm for superparticles, we are now ready to
explore the effect of particle collisions on particle concentration by
streaming instabilities and planetesimal formation by self-gravity. The
streaming instability feeds off the relative (streaming) motion of gas and
particles in protoplanetary discs and has a characteristic length scale
comparable to the sub-Keplerian length $\eta r$ \citep{YoudinGoodman2005}. Here
$\eta$ is the radial pressure gradient parameter of \cite{Nakagawa+etal1986}
and $r$ is the distance to the central star. \cite{Johansen+etal2009} and
\cite{BaiStone2010b} demonstrated that the streaming instability leads to
strong particle clumping when the heavy element abundance of the disc is above
a threshold value of $Z\approx0.02$ for particle sizes $\varOmega \tau_{\rm f}
\gtrsim 0.1$ \citep[and moderate radial drift, see][]{BaiStone2010c}. Clumping
proceeds as initially very low amplitude particle overdensities accelerate the
gas towards the Keplerian speed, hence reducing the local head-wind, which in
turn slows the radial drift of the particles. Drifting particles pile up where
the head-wind is slower, causing exponential growth of the particle density as
the particles continue to increase their drag force influence on the gas.
\cite{Johansen+etal2009} found that overdense regions contract when including
particle self-gravity and that eventually a number of gravitationally bound
clumps form. These models nevertheless did not include any particle collisions.
\begin{figure}[!t]
  \includegraphics[width=8.7cm]{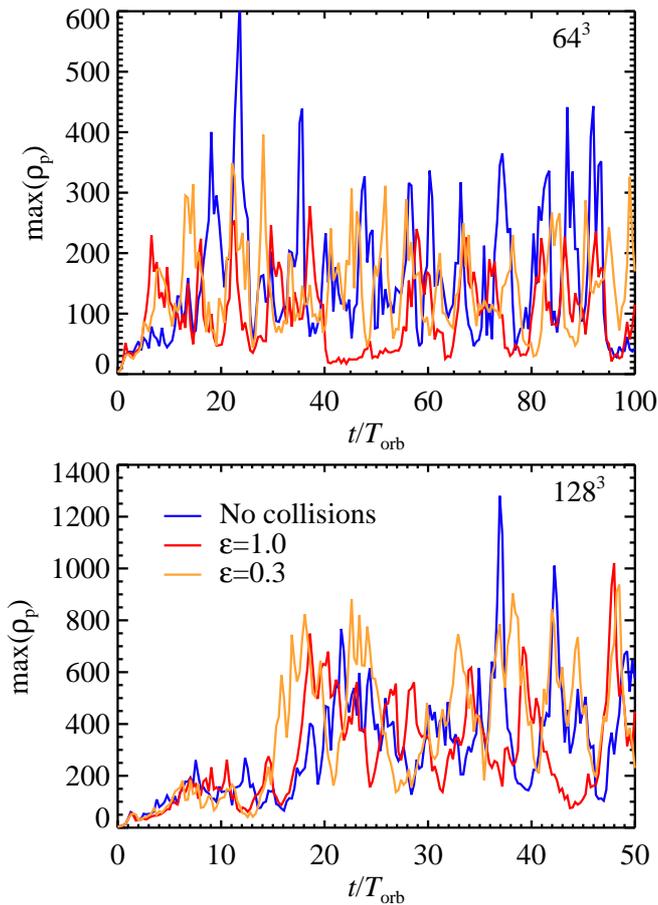}
  \caption{Maximum particle density, relative to the mid-plane gas density,
  versus time for a series of $64^3$ simulations (top plot) and $128^3$
  simulations (bottom plot) of turbulence driven by the streaming instability
  with different treatment of collisions. The maximum particle density
  increases by a factor approximately 2 when doubling the resolution, but the
  maximum density peaks are consistently 50\% lower when including particle
  collisions. Note the different scale of the axes in the two plots.}
  \label{f:rhopmax_t}
\end{figure}
\begin{figure}[!t]
  \includegraphics[width=8.7cm]{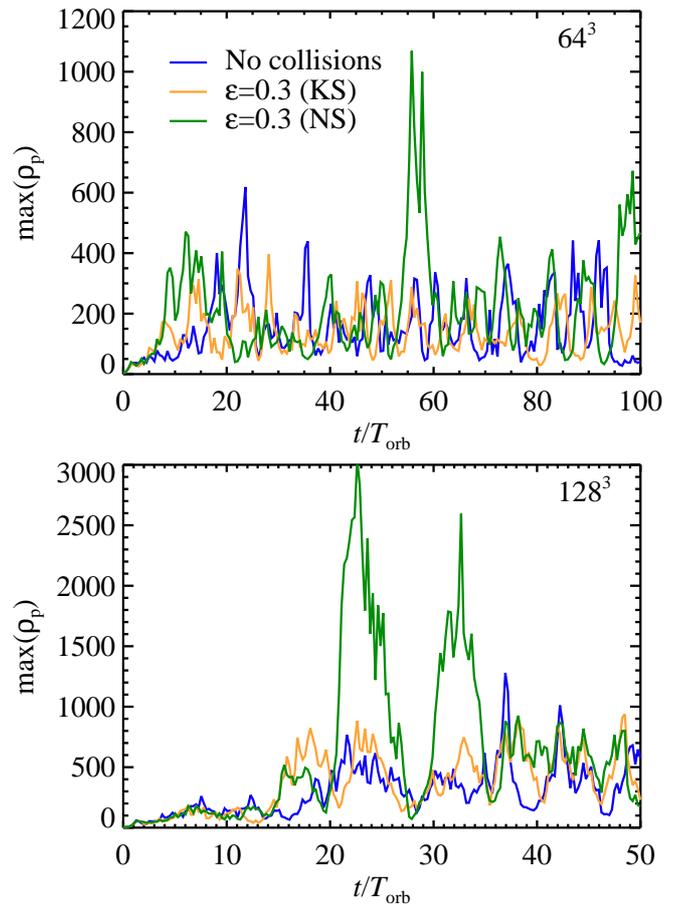}
  \caption{Maximum particle density, relative to the mid-plane gas density,
  versus time for simulations with normal collisions (KS) compared to
  simulations where we subtract off the Keplerian shear difference between
  particle pairs when calculating the collision time and the outcome of the
  collision (NS). NS collisions display more than three times higher particle
  densities than KS collisions. Peak concentrations fill a larger fraction of
  the simulation time at $128^3$.}
  \label{f:rhopmax_t_noshear}
\end{figure}
\begin{figure}[!t]
  \includegraphics[width=8.7cm]{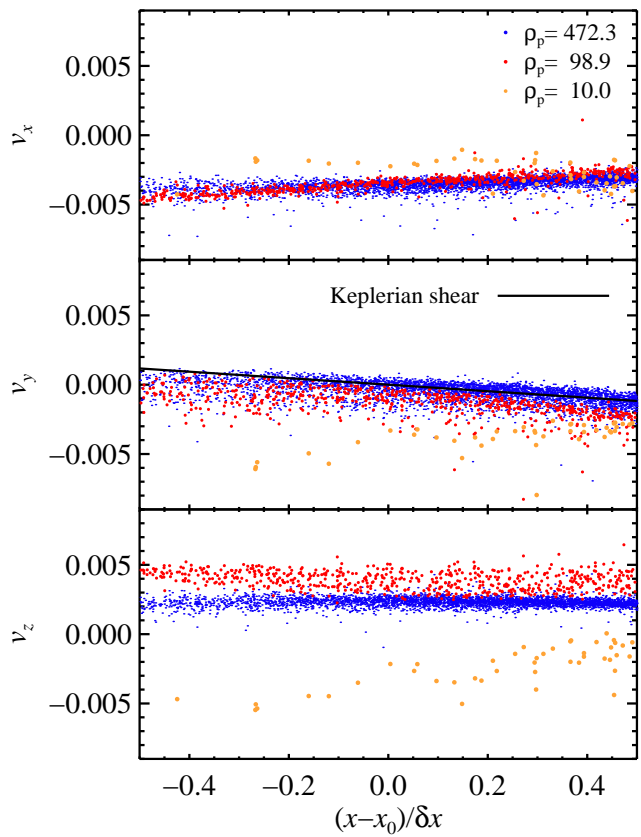}
  \caption{The three components of the particle velocity as a function of the
  radial position within a grid cell. Three grid cells were chosen at $t=45
  T_{\rm orb}$ of the run SI128\_e0.3, one with the highest particle density in
  the box, one with a particle density close to 100 times the gas density and
  finally one with a particle density close to 10 times the gas density. Both
  systematic and random particle motion is present within the grid cells. The
  Keplerian shear is clearly visible in the $y$-velocity (marked with a solid
  line in the middle panel). The cells with the highest density have generally
  a slower random motion and are thus more affected by the Keplerian shear.}
  \label{f:vp_gridcell}
\end{figure}

We perform 3-D simulations where the gas is modelled on a fixed grid and solid
particles with superparticles. We solve the standard shearing box equations for
gas and particles \citep[same as in][but with additional vertical
gravity]{JohansenYoudin2007}. The frame rotates at the Keplerian frequency
$\varOmega$ at a fixed orbital distance $r$ from the star. The coordinate axes
are oriented such that $x$ points radially outwards, $y$ points along the
rotation direction of the disc, while $z$ points perpendicular to the disc
along $\vc{\varOmega}$. The gas is subjected to a radial pressure gradient
which reduces its orbital speed by the positive amount $\Delta v=0.05 c_{\rm
s}$. Particles do not feel this radial pressure gradient, and the resulting
relative motion between particles and gas drives the streaming instability
\citep{GoodmanPindor2000,YoudinGoodman2005}. We consider a cubic box with side
lengths $L_x=L_y=L_z=0.2 H$, where $H=c_{\rm s}/\varOmega$ is the gas scale
height, to capture the fastest growing modes of the streaming instability of
marginally coupled particles, $\lambda_{\rm SI}/H\sim\eta r/H\sim\Delta
v/c_{\rm s}=0.05$. This is also the characteristic scale of Kelvin-Helmholtz
instabilities, thriving in the vertical shear in the gas and particle velocity
\citep{YoudinShu2002,Lee+etal2010}, although \cite{BaiStone2010b} demonstrated
that the streaming instability is dominant over Kelvin-Helmholtz instabilities
in setting the dynamics of particle layers with $\varOmega \tau_{\rm f}>0.1$.

The friction time of the particles is fixed at $\varOmega \tau_{\rm f}=0.3$ in
all simulations, corresponding to approximately 20-cm rocks around the location
of the asteroid belt at 3 AU, and to 6-mm pebbles at 30 AU
\citep{Weidenschilling1977}. The particle column density is set to 2\% of the
total gas column density, the latter including the gas beyond the vertical
boundaries of the box. For our choice of $\Delta v$ strong particle clumping
can only be obtained at such super-solar metallicity\footnote{The threshold for
clumping can be estimated analytically to be $Z\sim\eta (r/H)$
\citep{YoudinShu2002}. \cite{BaiStone2010c} and \cite{Johansen+etal2007}
confirmed numerically that the threshold for particle clumping by the streaming
instability shifts towards higher (lower) metallicity as the sub-Keplerian
speed difference $\Delta v$ is increased (decreased).}. The average dust-to-gas
ratio in a box of $L_z=0.2 H$ is $\langle \rho_{\rm p}/\rho_{\rm g}\rangle
\approx 0.25$ when $Z=0.02$. We set sound speed $c_{\rm s}$, Keplerian
frequency $\varOmega$ and mid-plane gas density $\rho_0$ to unity, so these
form the natural units of the simulations.

We compare results obtained without and with particle collisions. Simulations
with particle collisions are run in three variations: either with elastic
collisions ($\epsilon=1.0$), with inelastic collisions ($\epsilon=0.3$) or with
inelastic collisions where Keplerian shear is subtracted off when determining
the time-scale and outcome of collisions. Simulation parameters are given Table
\ref{t:parameters}. Each particle swarm contains a mass per volume of
$\hat{\rho}_{\rm p}/\rho_0\approx0.219$ for the considered particle number at
both $64^3$ and $128^3$.

\subsection{Maximum particle density}

We monitor the maximum particle density regularly in the simulations. In
\Fig{f:rhopmax_t} we show the maximum particle density versus time in
simulations with $64^3$ grid cells and $128^3$ grid cells, respectively.
Simulations without collisions generally achieve higher particle density -- up
to 600 times the gas density at $64^3$ and 1200 times the gas density at
$128^3$. Elastic collisions and inelastic collisions with $\epsilon=0.3$ give
very high particle densities too, but the peaks have an approximately 50\%
lower value than in simulations without collisions. Elastic collisions achieve
a somewhat lower maximum density than inelastic collisions. The kinetic energy
dissipation in inelastic collisions reduces the random motion of the particles
and allows higher particle contraction.

The inclusion of Keplerian shear during the collision can lead to unphysical
results, since the shear term is exaggerated by enlarging particles to the size
of a grid cell. The exaggerated kinetic energy input will in turn suppress
concentration peaks, in agreement with what is seen in \Fig{f:rhopmax_t}. In
\Fig{f:rhopmax_t_noshear} we show the maximum density in simulations with
inelastic KS and NS collisions respectively (and the results without collisions
for comparison). Simulations with NS collisions display a three times higher
maximum density than simulations with KS collisions. The maximum density is
even a factor 2-3 times higher than in simulations without collisions. This way
collisions actually promote particle concentration.

In \Fig{f:vp_gridcell} we analyse the particle motion within three grid cells
of the run SI\_128\_e0.3. We choose the grid cell with the maximum particle
density in the box and two grid cells with a particle density close to 100 and
10 times the gas density, respectively. The particle velocity shows both
systematic trends and random motion within the cells. The random motion is
slower in the cells of higher density. The Keplerian shear is clearly visible
in the $y$-velocity of particles in the two densest grid cells. Thus the
hydrodynamical simulations are prone to spurious heating, as explained above.
Subtracting off the Keplerian shear term when determining the time-scale and
outcome of collisions avoids this problem. \Fig{f:vp_gridcell} also shows a
systematic trend in the radial particle velocity. Radial convergence and
divergence in the particle velocity are expected when particles concentrate in
radial bands and when the concentrations dissolve again. We do not attempt to
correct for this systematic velocity within grid cells, but note that
systematic trends from smooth gradients will decrease with increasing
resolution.

\subsection{Particle concentration versus scale}

Overdense particle sheets contract radially under the action of self-gravity
and drag forces \citep{Youdin2011,Michikoshi+etal2010,ShariffCuzzi2011}. A full
non-axisymmetric collapse is initiated when the particle density crosses the
Roche density
\begin{equation}
  \rho_{\rm R} = \frac{9}{4 \pi} \frac{\varOmega^2}{G} \, .
\end{equation}
The mass of the planetesimal will be characterized by the scale over which the
Roche density is achieved. To quantify the scale-dependence of the particle
concentrations, we measure the maximum particle density over cubic regions of
side length $N_{\rm t}$ grid cells, increasing $N_{\rm t}$ from 1 to $N_x$. We
ensure that all concentrations centres are probed by stepping the measurement
region through the entire grid. Measurement regions crossing the boundaries are
handled by expanding the particle density field with its periodic counterpart
in all directions (glueing together $3^3$ copies which are identical except for
a shift due to Keplerian shear).
\begin{figure}[!t]
  \includegraphics[width=8.7cm]{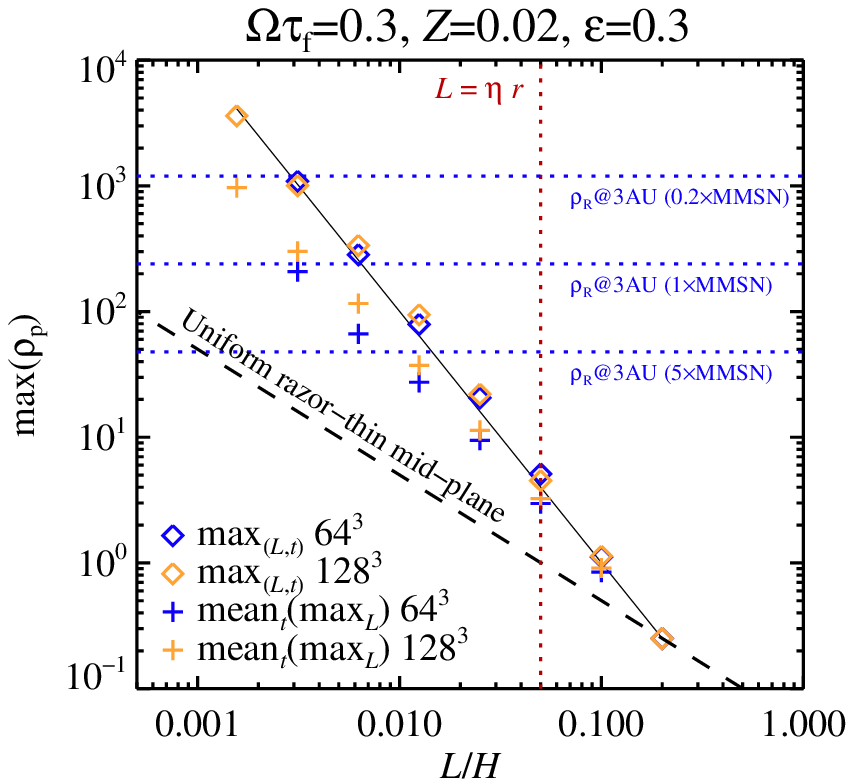}\\
  \includegraphics[width=8.7cm]{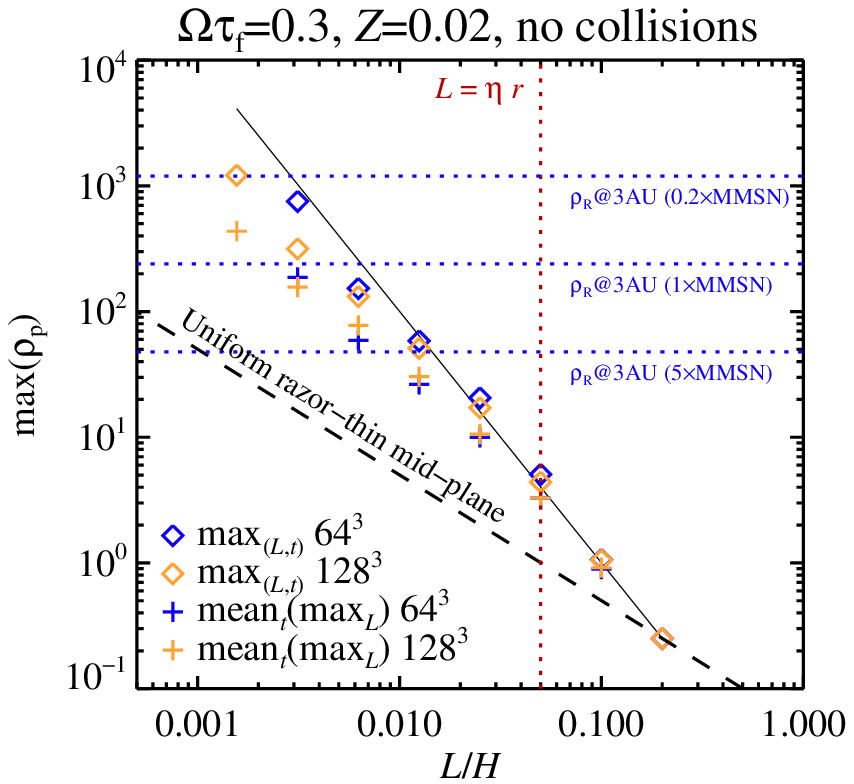}
  \caption{Maximum particle density, relative to the mid-plane gas density, as
  a function of scale, for simulations with NS collisions (top panel) and
  simulations with no collisions (bottom panel). Diamonds indicate the maximum
  density over a given scale, while pluses indicate the mean of the
  time-dependent maximum density. Simulations with NS collisions display good
  convergence in the maximum density, following closely a ${\rm max}(\rho_{\rm
  p})\propto L^{-2}$ law (thin black line), while the mean of the maximum
  density increases from $64^3$ to $128^3$, due to a higher temporal filling
  factor of major concentration events at higher resolution (see
  \Fig{f:rhopmax_t_noshear}). The dashed line shows the maximum density for a
  uniform razor-thin mid-plane layer for comparison. Blue dotted lines show the
  Roche density for the minimum mass solar nebula at 3 AU from the central
  star, and for five times less and more massive nebulae. The red dotted line
  indicates the characteristic length scale of the streaming instability,
  $L=\eta r$. Particle densities above $10^3$ times the gas density are reached
  in regions smaller than $\approx 0.003 H$, equivalent of $L \approx 50{,}000$
  km at 3 AU.}
  \label{f:rhopmax_scale}
\end{figure}

For snapshots saved once per orbit from $t=20T_{\rm orb}$ to $t=50T_{\rm orb}$
we calculate the maximum particle density as a function of scale. The results
are shown in \Fig{f:rhopmax_scale} for simulations with NS collisions
(SI64\_e0.3\_NS and SI128\_e0.3\_NS) in the top panel and simulations with no
collisions (SI64\_nocoll and SI128\_nocoll) in the bottom panel. We extend the
measurements of SI64\_e0.3\_NS to $t=60T_{\rm orb}$ to catch a major
concentration event (see top panel of \Fig{f:rhopmax_t_noshear}). We indicate
in \Fig{f:rhopmax_scale} both the maximum density over all times and the mean
of the time-dependent maximum density. The maximum scale-dependent density in
NS simulations is very similar at $64^3$ and at $128^3$. This quantity is
nevertheless very sensitive to the low-number statistics of the concentration
events. A more robust measure is the mean of the maximum density. This measure
increases somewhat from $64^3$ to $128^3$. It is also evident from
\Fig{f:rhopmax_t_noshear} that major concentration events have a higher
temporal filling factor at $128^3$. Whether this is intrinsic to the streaming
instability dynamics or just an effect of running simulations for too short
time is not possible to discern.
\begin{figure}[!t]
  \includegraphics[width=8.7cm]{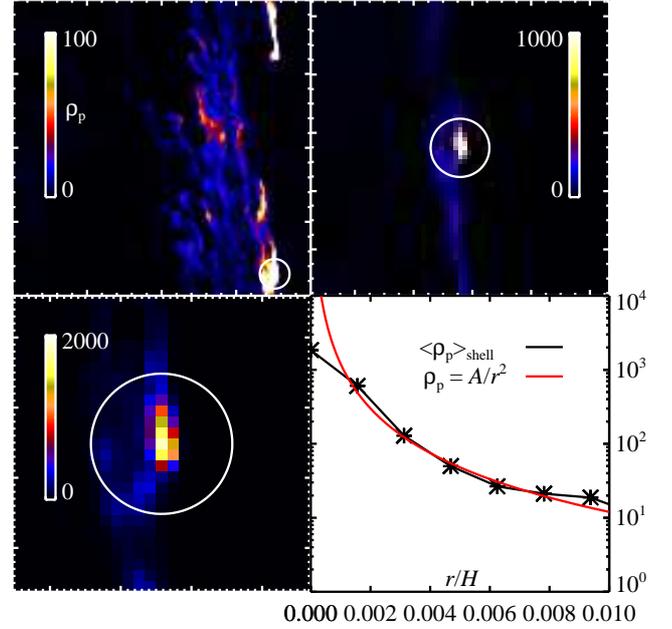}
  \caption{Zoom in on the densest grid cell in SI128\_e0.3\_NS at $t=32 T_{\rm
  orb}$. The overdense particle structure is elongated along the shear
  direction with a density decreasing in all directions from the densest
  point. The lower-right panel shows the particle density average over shells
  of thickness one grid cell and a $1/r^2$ power-law overplotted.}
  \label{f:rhop_spherical_clump}
\end{figure}

The apparent linear decrease of logarithmic density with logarithmic scale
implies ${\rm max}(\rho_{\rm p}) \propto L^{-\alpha}$ as a good model for the
scale-dependence of the maximum density. Two limits can immediately be put on
$\alpha$. The lowest value would stem from a razor-thin particle mid-plane
layer of uniform density, with $M \propto L^2$, giving ${\rm max}(\rho_{\rm p})
\propto M/L^3 \propto L^{-1}$ and thus $\alpha=1$. Concentration of all
particles in a single point would yield the upper limit of $\alpha=3$. We
overplot in \Fig{f:rhopmax_scale} with a thin black line the power law ${\rm
max}(\rho_{\rm p}) \propto L^{-2}$, fitted to match the mean density of the box
at $L=0.2 H$. The $\alpha=2$ power law follows the data extremely well. This
implies that $M \propto L$, i.e.\ that the particles primarily concentrate
either in 1-D filaments or in spherically symmetric clouds of density $\rho(r)
\propto 1/r^2$, known in star formation as the singular isothermal sphere
solution \citep[e.g.][]{Shu1977}. In \Fig{f:rhop_spherical_clump} we show the
particle density around the densest grid point in SI128\_e0.3\_NS at $t=32
T_{\rm orb}$. The overdense structure appears elongated along the $y$-direction
with the density falling rapidly towards all directions (although slower along
$y$).

Simulations without collisions (bottom panel of \Fig{f:rhopmax_scale}) show
similar trends as the simulations with NS collisions, but there is a marked
decrease in the maximum density over the smallest shared scale between $64^3$
and $128^3$. Nevertheless the mean of the maximum density agrees between the
two resolutions.

The convergence in scale-dependent maximum density shows that the dynamics of
the streaming instability concentration events is well-resolved and independent
of dissipation scale and viscosity. This is in contrast to turbulent
concentration in driven isotropic turbulence which, for a given particle size,
appears on length scales that are fixed relative to the Kolmogorov (viscous)
scale \citep{HoganCuzzi2007,Pan+etal2011}. In contrast the streaming
instability is fixed relative to the sub-Keplerian scale $\eta r\sim0.05 H$.
At $\ell \sim 0.0016 H$, probed only at $128^3$, the maximum density in
simulations with NS collisions reaches more than three thousand times the gas
density. Higher resolution simulations will be needed to test if the particle
density continues to follow the ${\rm max}(\rho_{\rm p})\propto L^{-2}$ trend,
or eventually finds a smallest scale. The 2-D streaming instability simulations
of \cite{BaiStone2010a} converged in density statistics at between $512^2$ and
$1024^2$ grid cells. Reaching those resolutions in 3-D is very computationally
demanding, but should be an important priority for the future.
\begin{figure*}
  \includegraphics[width=17.3cm]{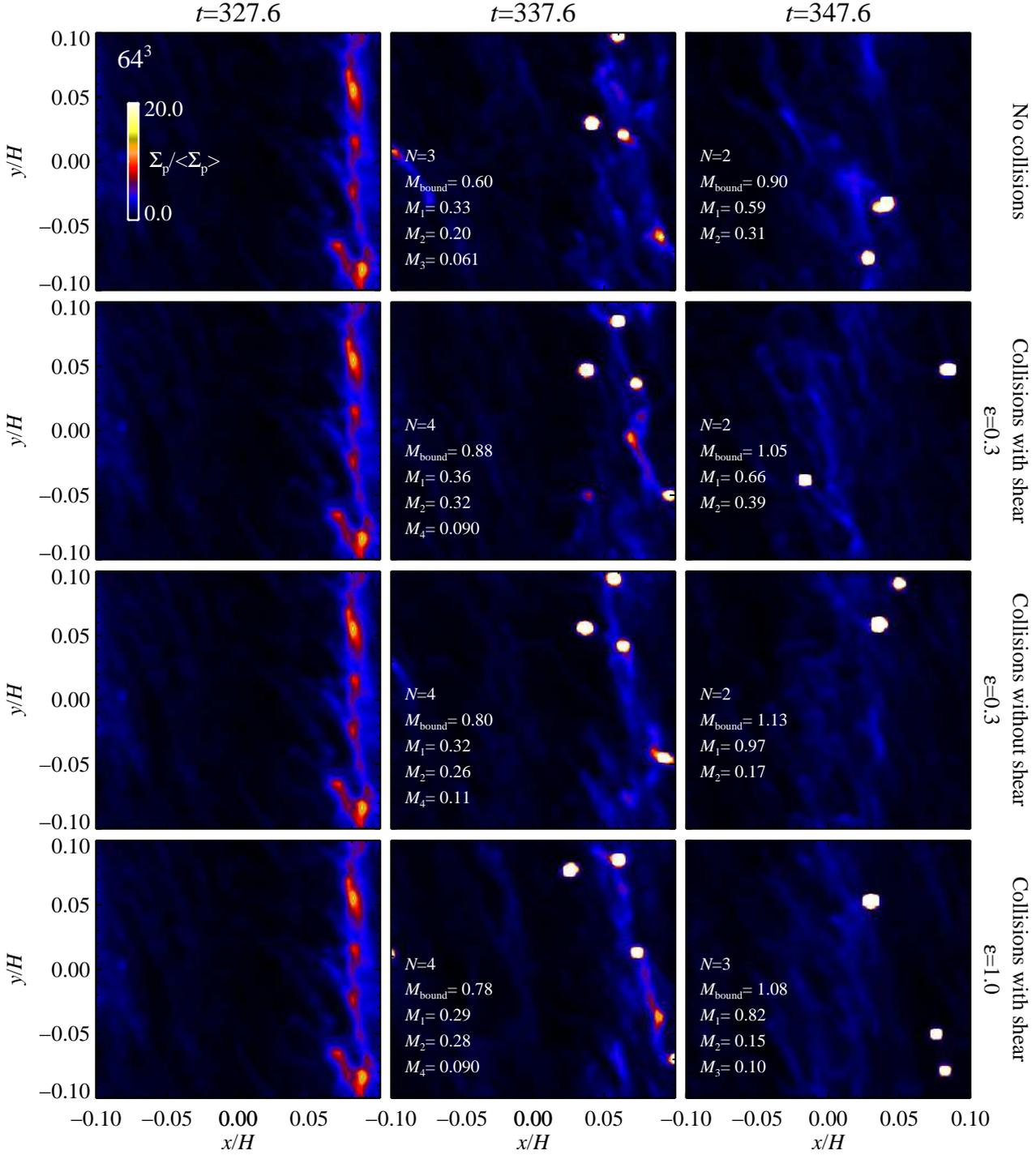}
  \caption{Particle column density versus time after self-gravity is turned on
  at $t_0=52 T_{\rm orb} = 326.726 \varOmega^{-1}$ in the simulation
  SI64\_e0.3\_NS. An overdense sheet forms by the streaming instability and
  breaks up in a number of gravitationally bound clumps. We indicate the number
  of clumps and their masses, in units of the mass of the dwarf planet Ceres,
  in the lower left part of the plots. Between 3 and 4 clumps condense out
  independently of how collisions are treated, with masses slightly smaller
  than Ceres. Clump merging, likely driven by the artificially large sizes of
  the planetesimals, reduces the number of clumps with time in all cases. Note
  that the initial condition for all four simulations is taken from
  SI64\_e0.3\_NS.}
  \label{f:sigmap_t_panels_64}
\end{figure*}
\begin{figure*}
  \includegraphics[width=17.3cm]{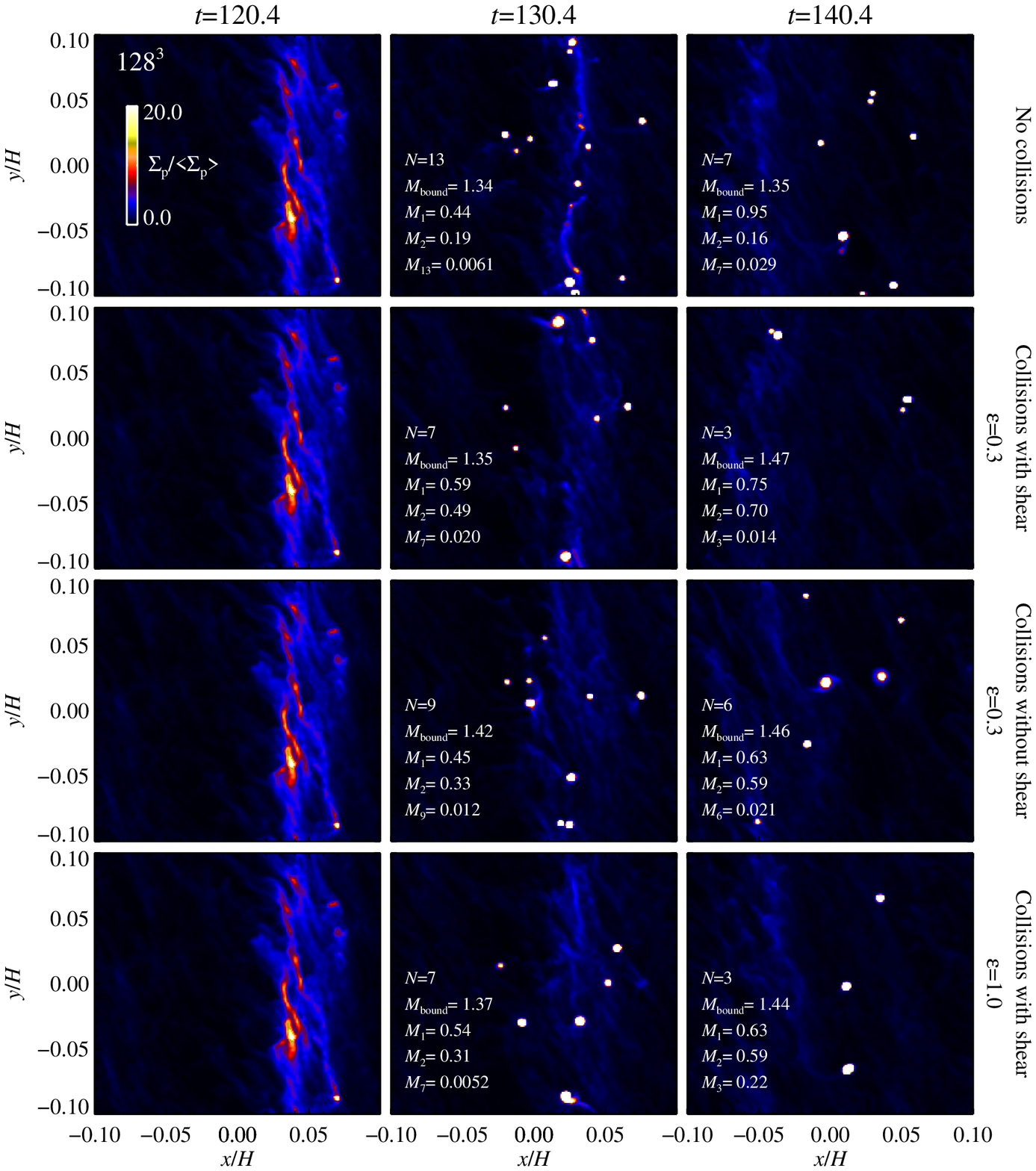}
  \caption{Same as \Fig{f:sigmap_t_panels_64}, but for $128^3$ simulations with
  self-gravity started at $t_0=19 T_{\rm orb} = 119.381 \varOmega^{-1}$. More
  clumps form initially, but the most massive clumps have similar masses to the
  $64^3$ simulation. The run with no collisions forms more low-mass clumps than
  the other runs. The initial condition for all four simulations is taken from
  SI128\_e0.3\_NS. The total particle mass in the box is approximately 2.8
  Ceres masses.}
  \label{f:sigmap_t_panels_128}
\end{figure*}

\section{Planetesimal formation}
\label{s:planetesimals}

The gravitational potential field of the particles is found by mapping the
particle density on the grid, using a second order spline interpolation scheme,
and solving the Poisson equation using a fast Fourier transform method
\citep{Johansen+etal2007}. The gravitational acceleration is interpolated back
to the particle positions using second order spline interpolation. The strength
of the gravity is defined by the non-dimensional parameter
\begin{equation}
  \tilde{G}=\frac{4 \pi G \rho_0}{\varOmega^2} \, ,
\end{equation}
which is related to the thin-disc self-gravity parameter $Q$ through $Q \approx
1.6 \tilde{G}^{-1}$ \citep{Safronov1960,Toomre1964}. The solar nebula of
\cite{Hayashi1981} has $\tilde{G}\approx0.04$ at 3 AU from the sun, the
parameter depending weakly on the distance. We use $\tilde{G}=0.1$ as a
reference choice in the simulations, but experiment with $\tilde{G}$ down to
0.02.

The total particle mass in the box is
\begin{equation}
  M_{\rm p} = \langle \rho_{\rm p} \rangle L^3 \approx 0.002 H^3 \rho_0 \, ,
  \label{eq:munit}
\end{equation}
where the mass unit $M_0 = H^3 \rho_0$ depends on the temperature and location
in the disc [$H$] and the strength of the self-gravity [$\rho_0=(4 \pi G)^{-1}
\tilde{G} \varOmega^2$]. While the expression in \Eq{eq:munit} does not depend
on $\tilde{G}$, in units where $H=\rho_0=1$, the {\it physical} mass unit does.
In a nebula with the scale-height given by \cite{Hayashi1981}, we have at
$r=3\,{\rm AU}$ with $\tilde{G}=0.1$ a mass unit of $M_0 \approx 1.3 \times
10^{27}\,{\rm g}$ and $M_{\rm p} \approx 2.8 M_{\rm Ceres}$.

We activate particle self-gravity in simulations of the streaming instability
with inelastic NS collisions, at times when there is little particle
concentration, to catch the simultaneous action of streaming instability and
self-gravity during the next concentration event. In SI64\_e0.3\_NS we thus
start self-gravity at $t=52T_{\rm orb}$, while in SI128\_e0.3\_NS we start
self-gravity at $t=19 T_{\rm orb}$ (see \Fig{f:rhopmax_t_noshear}). We then
evolve the simulation for another 5 orbits, either ignoring collisions or
applying the usual variation of collision types (elastic, inelastic KS,
inelastic NS).

Results of $64^3$ simulations are shown in \Fig{f:sigmap_t_panels_64}. Between
3 and 4 clumps\footnote{The algorithm for identifying bound clumps is based on
2-D column density snapshots and is described in detail in
\cite{Johansen+etal2011}.} initially condense out of the dominantly
axisymmetric filament forming by the streaming instability. These clumps have
masses between a tenth and a third of the dwarf planet Ceres -- corresponding
to contracted radii between 220 and 330 km, assuming an internal density of 2
g/cm$^3$. All the clumps form in a single planetesimal-formation event shortly
after the onset of self-gravity. The clumps continue to grow mainly by
accreting particles from the turbulent flow, but no new gravitationally bound
clumps form. Clumps eventually collide and merge in all simulations. Such clump
merging is likely an unphysical effect driven by the large sizes of the
planetesimals. The self-gravity solver does not allow gravitational structures
to become smaller than a grid cell, and that leads to artificially large
collisional cross sections. A more probable outcome of the real physical system
is gravitational scattering and/or formation of binaries
\citep{Nesvorny+etal2010}.

Results at $128^3$ are shown in \Fig{f:sigmap_t_panels_128}. At higher
resolution the number of clumps condensing out is about twice as high compared
to the lower resolution simulation. However, the masses of the most massive
clumps are very similar to lower resolution (although a bit higher -- up to
60\% of Ceres), so it appears that higher resolution simply allows lower-mass
clumps to condense out as well. The masses of the clumps condensing out at
$128^3$ resolution correspond to contracted radii between 84 and 405 km. The
ability to form smaller clumps at higher resolution is expected from the
picture that a radial contraction phase is needed before the Roche density can
be achieved\footnote{A similar order of events is seen in simulations of star
formation in self-gravitating accretion discs around supermassive black holes,
see e.g.\ Fig.\ 3 of \cite{Alexander+etal2008}.}. Higher resolution allows
contraction to narrower bands and thus formation of less massive
planetesimals. It is nevertheless difficult to compare the planetesimal masses
condensing out at the two resolutions as the initial conditions are not the
same.

\cite{Rein+etal2010} observed in their 2-D shearing sheet simulations that
inclusion of collisions would lead to condensation of fewer and more massive
clumps, when compared to simulations without collisions. Our
\Fig{f:sigmap_t_panels_128} also shows that the simulation with no collisions
makes the highest number of clumps of all the four simulations. Nevertheless
the characteristic mass of the most massive clumps appears indifferent to the
treatment of collisions.

\begin{figure}
  \includegraphics[width=8.7cm]{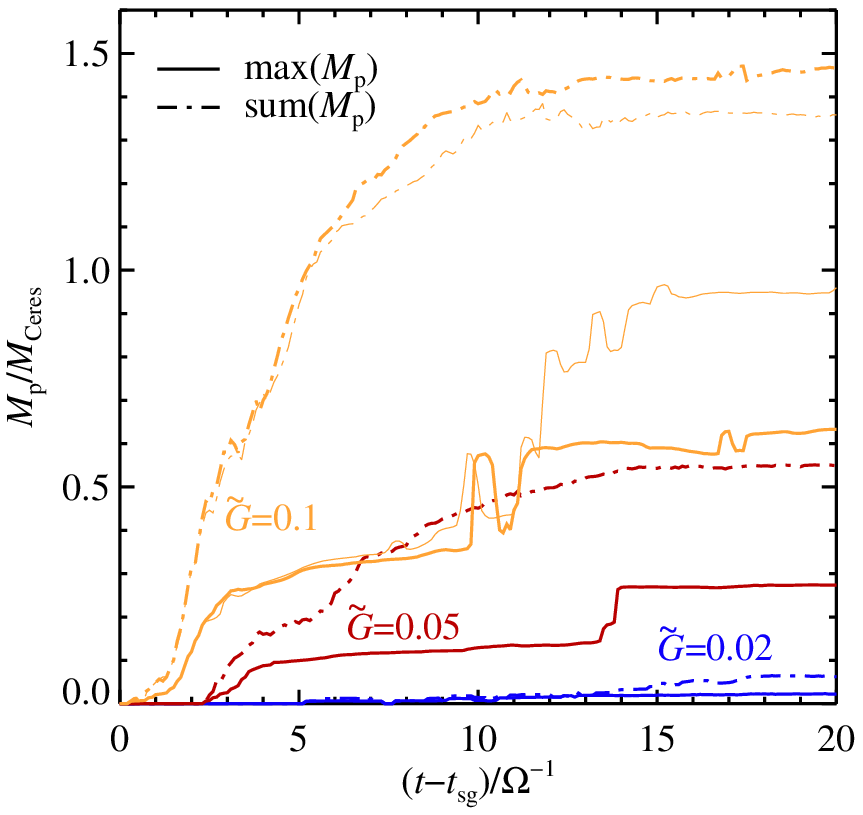}
  \caption{Evolution of maximum planetesimal mass (full line) and total mass in
  planetesimals (dash-dotted line) for $128^3$ simulations with inelastic NS
  collisions (thin yellow line shows the $\tilde{G}=0.1$ simulation without
  collisions for comparison). Colors indicate $\tilde{G}=0.02, 0.05, 0.1$.
  Extended wiggles in the $\tilde{G}=0.1$ curve arise during clump merging.
  The total particle mass in the box is $2.8$, $1.4$ and $0.56$ Ceres masses,
  in order of decreasing $\tilde{G}$.}
  \label{f:mpmax_t}
\end{figure}

Since $\tilde{G}$ controls the relative strength of self-gravity, results
obtained with a given $\tilde{G}$ can not be scaled to other values of
$\tilde{G}$. We vary the self-gravity parameter in $128^3$ simulations in
\Fig{f:mpmax_t}, starting self-gravity at the same time as in
\Fig{f:sigmap_t_panels_128}. Weaker self-gravity gives lower clump masses, but
gravitationally bound clumps of up to $0.01$ Ceres masses (or 100 km radius)
condense even at $\tilde{G}=0.02$. The solar nebula model of \cite{Hayashi1981}
has $\tilde{G}\approx0.04$ at 3 AU from the sun. Thus the streaming instability
allows planetesimal formation in disc models that are similar in mass to the
solar nebula, in contrast to recent simulations of planetesimal formation in
pressure bumps excited by the magnetorotational instability which required disc
masses up to 10 times the solar nebula \citep{Johansen+etal2011}.

The presented simulations do not catch the transition from bound clump to solid
planetesimal. However, \cite{Nesvorny+etal2010} simulated the gravitational
collapse of spherical particle clouds and generally found formation of binary
planetesimals, with the two largest bodies containing a significant fraction of
the mass of the cloud. The fact that the masses of the most massive bound
clumps in our simulations are relatively independent of resolution allows us to
critically compare the mass distribution of the clumps to to the observed
properties of the asteroid and Kuiper belts and extrasolar debris discs.

\subsection{Application to the Kuiper belt}

The physical mass of the clumps depends on location in the disc and on the
self-gravity parameter $\tilde{G}$. While the simulations are dimensionless,
the translation to physical mass involves multiplication by the mass unit $M_0
= \rho_0 H^3 = \tilde{G} \varOmega^2 H^3/(4 \pi G) $. In a nebula with constant
$\tilde{G}$ and $T\propto r^{-1/2}$, the mass unit scales as $M_0 \propto
r^{3/4}$, so re-scaling to the Kuiper belt\footnote{The orbits of
trans-Neptunian objects extend to several 10 AU beyond the orbit of Neptune,
although many of these must have formed within the orbit of Neptune and been
scattered outwards later. Thus we take 30 AU as an approximate distance scale.}
gives planetesimal masses 5--6 times higher than in \Fig{f:sigmap_t_panels_64}
and \Fig{f:sigmap_t_panels_128}. Contracted radii at the location of the Kuiper
belt are approximately 80\% higher than in the asteroid belt, yielding
planetesimal radii between 150 and 730 km. The upper range is comparable to the
masses of the largest known Kuiper belt objects
\citep{Chiang+etal2007,Brown2008}.

This extrapolation is only valid for an assumed constant self-gravity parameter
$\tilde{G}$. The minimum mass solar nebula, with $\varSigma\propto r^{-3/2}$,
has $\tilde{G}\propto r^{1/4}$. The weak dependence on radial distance from the
star gives in the Kuiper belt at $r=30\,{\rm AU}$ a $10^{1/4}\approx1.8$ times
larger $\tilde{G}$ than in the asteroid belt. From \Fig{f:mpmax_t} we read off
an approximate doubling in planetesimal mass when increasing $\tilde{G}$ from
$0.05$ to $0.1$. We expect that this scaling holds for larger $\tilde{G}$ as
well. This way the minimum mass solar nebula gives somewhat higher masses in
the Kuiper belt compared to the constant-$\tilde{G}$ extrapolation presented
above.

The comparison to observed planetesimal belts is nevertheless complicated by a
potentially very efficient accretion of unbound particles (pebbles and rocks)
by the newly born planetesimals after their formation
\citep{JohansenLacerda2010,OrmelKlahr2010}, an epoch not captured in our
simulations. It is interesting to note that, given the power of the streaming
instability in producing Ceres-mass planetesimals from pebbles and rocks, the
challenging question may not be how these planetesimal belts form\footnote{This
does require sufficient amounts of pebbles and rocks to begin with, the
formation of which is not yet well-understood \citep{BlumWurm2008}.} or how the
characteristic mass arises, but rather why the planetesimals did not
immediately continue to grow towards terrestrial planets, super-Earths, and
cores of ice and gas giants. Perhaps these planetesimal bursts were
``abandoned'' by the particle overdensity from which they formed, by radial
drift of the particles, stranding as planetesimal belts. Such stranding is
evident in the last frames of \Fig{f:sigmap_t_panels_64} and
\Fig{f:sigmap_t_panels_128} where the gravitationally bound clumps clearly lag
behind the overdense particle filament. The lag might have be even more
pronounced if the particle clumps would not be bloated to fill a grid cell.

This stranding scenario is an alternative to the more classical view that the
asteroid and Kuiper belts were disturbed by the presence of giant planets
\citep[e.g.][]{KenyonBromley2004}.

\section{Summary and discussion}
\label{s:discussion}

This paper focuses on the effect of momentum exchange and energy dissipation in
collisions on particle concentration by the streaming instability and on the
subsequent gravitational collapse to form dense clumps and planetesimals. We
develop a new algorithm for tracking collisions between superparticles
representing swarms of physical particles. The time-scale for a particle in a
given swarm to collide with a particle from another swarm is calculated for all
superparticle pairs in a grid cell. Collisions occur instantaneously if a
random number is less than the ratio of the simulation time-step to the
collisional time-scale, ensuring that superparticles collide statistically on
the correct time-scale. We have demonstrated that this algorithm can be
incorporated into a hydrodynamical code at a modest computational cost. This is
true even for large particle numbers, since the number of possible collision
partners that are considered in a given timestep can be reduced with little or
no loss of generality.

Collisions can have a number of effects on particle dynamics, by making
particle motion more isotropic and by dissipative collisions which drain
kinetic energy from the system. We have considered the simplest case of a
constant coefficient of restitution (either unity or 0.3), but a more
physically motivated coefficient of restitution, depending on material
properties and impact speed and angle, could be easily implemented in the
scheme. We emphasize that we have focused in this paper entirely on particles
with a friction time of 0.3 relative to the local Keplerian time-scale,
corresponding to 20-cm rocks in the asteroid belt and 6-mm pebbles at 30 AU.
Future studies will be needed to determine the influence of particle collisions
on the dynamics of smaller and larger particles and on their ability to form
planetesimals.

Our simulations show that collisions are important to consider when modelling
particle concentration by the streaming instability. Taking into account the
energy dissipation in inelastic collisions increases the maximum particle
density. This increase is most pronounced, more than a factor of three compared
to simulations with no collisions, when we ignore the relative Keplerian shear
for determining the collision time-scales and outcomes. We argue that the
Keplerian shear velocity should be subtracted when determining the outcome of
collisions between superparticles representing physical particles that are much
smaller than a grid cell. The collision algorithm enlarges particles to the
size of a grid cell during a collision, and this can lead to unphysical heating
of the particle component if the Keplerian shear is included during the
collision.

The treatment of collisions has no apparent effect on the planetesimals which
form by self-gravity. The masses of the most massive planetesimals are
relatively independent of the inclusion or absence of collisions, although we
find some evidence that more low-mass clumps condense out in simulations
without collisions. The particle densities reach several hundred and even
thousand times the gas density both with and without collisions -- much higher
than the Roche density which governs gravitational collapse -- and that may
explain why particle collisions play a relatively small role in determining the
outcome of the gravitational contraction to form planetesimals. The simulations
show a characteristic planetesimal mass-scale comparable to the dwarf planet
Ceres at the location of the asteroid belt. The mass-scale increases
approximately linearly with distance from the central star, giving almost
double the contracted radius at the distance of the Kuiper belt. This scaling
may explain why the largest Kuiper belt objects are bigger than the largest
asteroids.

Particle collisions are also important as a stepping stone towards implementing
coagulation and fragmentation in planetesimal formation models
\citep{OrmelSpaans2008,ZsomDullemond2008}. Including all the physics relevant
for modelling particle-dominated self-gravitating flows is a major task, but
the reward will be a much better understanding of the important step from
pebbles and rocks to planetesimals and dwarf planets.

\begin{acknowledgements}

Y.L.\ acknowledges support from NSF grant AST-1109776.
Computer simulations were performed at the Platon system of the Lunarc Center
for Scientific and Technical Computing at Lund University. We are grateful to
Hanno Rein, Geoffroy Lesur, Zoe Leinhardt, Yuri Levin, Ross Church, Andras Zsom
and Kees Dullemond for stimulating discussions. We thank the referee, Chris
Ormel, for raising many interesting points in his very thorough referee report.
We thank the Isaac Newton Institute for Mathematical Sciences for providing an
environment for stimulating discussions during the ``Dynamics of Discs and
Planets'' programme.

\end{acknowledgements}

\appendix

\section{Collision time from friction time}
\label{s:taucfromtauf}

In connection with the presence of gas it is convenient to express the
collision time-scale in terms of the gas friction time-scale. In the Epstein
drag force regime, valid when the radius of a particle $R$ is smaller than
($9/4$ times) the mean free path of gas molecules \citep{Weidenschilling1977},
the friction time-scale is
\begin{equation}
  \tau_{\rm f} = \frac{R\rho_\bullet}{c_{\rm s}\rho_{\rm g}} \, .
\end{equation}
Here $\rho_\bullet$ is the material density of the particles, while $c_{\rm s}$
and $\rho_{\rm g}$ are the sound speed and density of the gas molecules.

The time-scale for a particle of radius $R_k$ to collide with a swarm of
particles with physical radius $R_j$ is
\begin{equation}
  \tau_{\rm c}^{(k)} = \frac{1}{\hat{n}_j \sigma_{jk} \delta v_{jk}} \, ,
  \label{eq:tcoll_nj}
\end{equation}
where $\sigma_{jk}$ is the mutual collisional cross section. Writing further
$\hat{n}_j=\hat{\rho}_j/m_j$ and $\sigma_{jk}=\pi(R_j+R_k)^2$ and assuming
spherical particles we arrive at
\begin{equation}
  \tau_{\rm c}^{(k)} = \frac{(4/3) \rho_\bullet R_j^3}{\hat{\rho}_j (R_j+R_k)^2
  \delta v_{jk}} \, .
\end{equation}
In terms of the friction time we get
\begin{equation}
  \tau_{\rm c}^{(k)} = \frac{4}{3}\tau_{\rm f}^{(j)} \frac{\rho_{\rm
  g}}{\hat{\rho}_j} \frac{c_{\rm s}}{\delta v_{jk}} \left( \frac{\tau_{\rm
  f}^{(j)}}{\tau_{\rm f}^{(j)}+\tau_{\rm f}^{(k)}} \right)^2 \, .
  \label{eq:tcoll_tfric}
\end{equation}
For collisions between equal-sized particles, with $\tau_{\rm
f}^{(j)}=\tau_{\rm f}^{(k)}$, the expression reduces to
\begin{equation}
  \tau_{\rm c} = \frac{\tau_{\rm f}}{3}
  \frac{\rho_{\rm g}}{\hat{\rho}_j} \frac{c_{\rm s}}{\delta v_{jk}} \, .
  \label{eq:tcoll_tfric_eq}
\end{equation}
A time-dependent numerical solution of a collisional particle system must take
collisions into account when choosing the time-step. The time-step criterion of
the Monte Carlo collision scheme originates in the requirement that two
particles can collide at most once during a time-step, i.e. the collision
probability $P=\delta t/\tau_{\rm c}$ between any two particles in the same
grid cell must be much smaller the unity. This time-step is independent of the
maximum density in a grid cell, since particles in dense grid cells have many
collision partners and hence can suffer more collisions in the same time-step.

In the streaming instability simulations presented in \Sec{s:streaming} and
\Sec{s:planetesimals} we observe a typical particle rms speed $\delta
v\sim0.025 c_{\rm s}$. The mass density represented by a single superparticle
is $\hat{\rho}_{\rm p} \approx 0.219 \rho_{\rm g}$ and the friction time is
$\varOmega \tau_{\rm f}=0.3$ (we normalise here by the Keplerian frequency
$\varOmega$ which we define in \Sec{s:streaming}). This gives $\varOmega
\tau_{\rm c} \approx 18$ from \Eq{eq:tcoll_tfric_eq}. The Courant criterion for
the hydrodynamical part of the streaming instability gives the time-step
$\delta t_{\rm hydro}=0.000625\varOmega^{-1}$ for $64^3$ and $\delta t_{\rm
hydro}=0.0003125\varOmega^{-1}$ for $128^3$ simulations. Therefore we can
ignore the collision time-scale in the simulations when determining the
numerical time-step.

\subsection{Multiple particle sizes}
\label{s:multiple}

\Eq{eq:tcoll_nj} defines the collisional time-scale between particles of two
sizes. For two superparticles of equal internal particle number ($\hat{n}$) we
have $\tau_{\rm c}^{(k)}=\tau_{\rm c}^{(j)}$, because the cross section
$\sigma_{jk}$ and relative speed $\delta v_{jk}$ are symmetric in $(j,k)$.
However, equal particle number per superparticle is numerically expensive, as
the mass of a superparticle in that case scales as $R^3$, requiring many more
superparticles to represent an equal mass of smaller particles. The second
complication is that the collision time-scale becomes very short for smaller
particles.

A more common approach is to have equal mass per superparticle. In that case we
can define a collision time-scale as the time for all mass in particle $j$ to
interact with all mass in particle $k$. This time-scale is shared between the
two particle species and is given by
\begin{equation}
  \tau_{\rm c} = \frac{4}{3}{\rm max}(\tau_{\rm f}^{(j)},\tau_{\rm
  f}^{(k)}) \frac{\rho_{\rm
  g}}{\hat{\rho_j}} \frac{c_{\rm s}}{\delta v_{jk}} \left( \frac{{\rm max}(\tau_{\rm
  f}^{(j)},\tau_{\rm f}^{(k)})}{\tau_{\rm f}^{(j)}+\tau_{\rm f}^{(k)}} \right)^2 \, .
  \label{eq:tcoll_tfric}
\end{equation}
To illustrate this, take small particles of friction time $\tau_{\rm
f}^{(j)}=1$ and large particles of friction time $\tau_{\rm f}^{(k)}=100$. The
collision time-scale for the large particles is 100 times shorter than for the
small particles, because the superparticle with small physical particles
contains 100 times more particles in the swarm. However, the time-scale for
collision between a large and a small particle does not imply that all small
particles have collided during that time. The correct time-scale is the time
for small particles to collide with large particles. When an average small
particle has experienced a collision, then all small particles have collided
with a large particle, and all the mass in the two superparticles have
interacted.

After waiting the common collision time $\tau_{\rm c}$, the collision outcome
can be solved as if the two colliding particles had equal mass, since
effectively a large particle collides with $m_k/m_j$ small particles during
this time. This approach is slightly inconsistent since discrete collisions
with $N$ particles of mass $m_j$ is not equal to a single collision with a
particle of mass $N m_j$. A collision between a particle of velocity $v_k$ and
a stationary particle results in the new velocity
\begin{equation}
  v_k' = \frac{m_k-m_j}{m_k+m_j} v_k \, .
\end{equation}
After $N$ such collisions the velocity of particle $k$ is
\begin{equation}
  v_k' = \left(\frac{m_k-m_j}{m_k+m_j}\right)^N v_k \, .
\end{equation}
In the limit where $v_k - v_k' = \Delta v_k \ll v_k$, this equation describes a
velocity damping
\begin{equation}
  \frac{\de v_k}{\de t} = -\frac{1}{\tau_{\rm c}} \frac{2 m_j}{m_k+m_j} v_k
\end{equation}
with characteristic time-scale $\tau_{\rm d}=\tau_{\rm c} (m_k+m_j)/(2 m_j)$.
Completely braking down the large particle requires infinite time, whereas a
single discrete collision with an equal-mass particle would remove all the
momentum from particle $j$ in one collision time.

To really get the collisional energy equipartition right between particles of
different sizes one would have to allow for collisions between a large particle
and individual smaller particles. This could either be done by letting
superparticles not represent the same mass, but rather the same number of
particles. However, this approach would become unpractical to model a large
span in particle sizes, since a huge number of superparticles would be required
to represent the low-mass particle bins. Alternatively the collision between a
swarm of large and small particles could be modelled on the collision
time-scale of individual collisions, distributing afterwards the energy and
momentum of the particle that suffered the collision among the entire swarm of
small particles or among all particle swarms within its mean free path.
However, for a large span in particle sizes, this would still require a very
small time-step and is therefore unpractical. We simply note here that while
collisions between unequal-sized particles can be modelled with the right
conservation properties, actual equipartition of particle energies would
require an adaptation of the collision algorithm.

\section{Limiting the collision number}
\label{s:limiter}

During the gravitational contraction of particle clumps the number of particles
in a grid cell can become very large, on the order of 1000s of 10000s. Tracking
$(1/2)N(N-1)$ possible collisions per grid cell then becomes very
computationally expensive.

However, particles do not collide with all possible partners during a single
time-step. One can limit the number of collision partners, while maintaining
the overall collision rate, by sampling only a subset $N_{\rm max}$ of the
possible collisions. Considering only $N_{\rm max}$ out of the $N-1$ collision
partners for each particle in a grid cell, while increasing the collision
probability for each collision partner by $(N-1)/N_{\rm max}$, yields
statistically the same number of collisions.

Consider as an example 101 particles in a grid cell, with the collision
probability between any two particles of $10^{-2}$. Particle 1 will then on the
average collide with 1 other particle. However, calculating the collision
probability with 100 other particles is expensive, even when it does not lead
to a collision, which is most often the case. Instead we let particle 1 only
interact with particles $2$ to $6$, and give each collision the probability
$10^{-1}$ instead of $10^{-2}$. Particle $i$ has particles $i+1$ to $i+5$ as
collision partners. When reaching particle $97$, the collision partners wrap
around to particle 1 again, and this way all particles on the average get 10
collision partners (5 of higher index and 5 of lower index) instead of 100.
\begin{figure}
  \includegraphics[width=8.7cm]{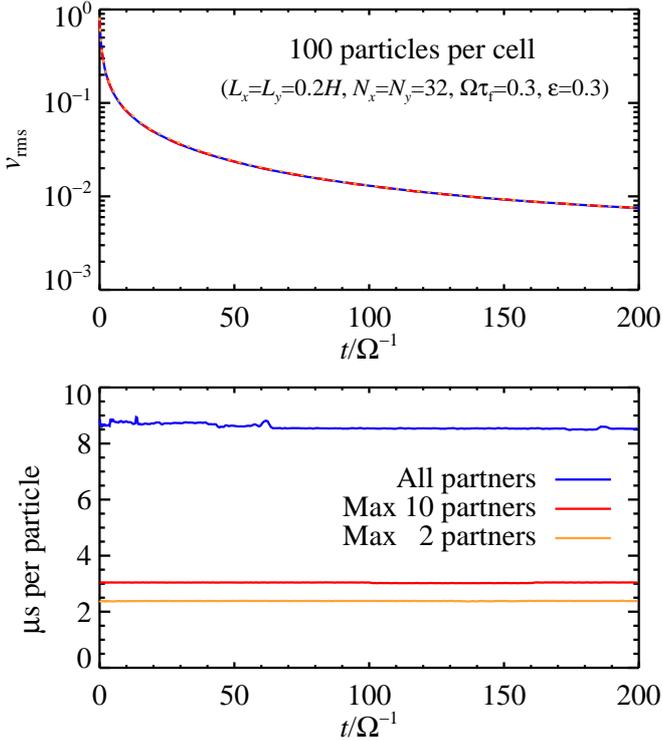}
  \caption{Evolution of particle rms speed in simulation starting with random
  motion of amplitude 1. Particles have mean-free-path $\lambda=0.1$ and
  coefficient of restitution $\epsilon=0.3$. Drag forces are ignored. The blue
  line shows the results of a simulation with 100 particles per grid cell and
  full collision partner list, while the red and golden lines show the results
  of limiting the collision partners to 10 and 2, respectively, while
  increasing the collision probability accordingly. The results are extremely
  similar. The lower panel shows the instantaneous inverse code speed. Limiting
  the number of collision partners has increased the speed by a factor of
  approximately three.}
  \label{f:vprms_t_npart_comparison}
\end{figure}

When reducing the number of collision partners, one has to be careful that the
particles do not interact only with particles of a nearby index in each
time-step. To avoid any such spurious particle preferences, we therefore
shuffle the order of particles inside a grid cell in each time-step. We have
empirically found that reducing the number of collision partners becomes
important when there are more than 100 particles in a grid cell. We show in
\Fig{f:vprms_t_npart_comparison} the rms speed of particles undergoing
inelastic collisions with coefficient of restitution $\epsilon=0.3$. We use 100
particles per grid cell and show results where we consider all particles in a
cell to be collision partners together with results where we limit the
collision partners to 10 and 2. The results are indistinguishable, but the code
speed is significantly higher when limiting the number of collision partners
(lower panel of \Fig{f:vprms_t_npart_comparison}). The typical speed of the
Pencil Code for a hydrodynamical simulation with two-way drag forces between
gas and particles is $\sim$10 $\mu$s per particle per time-step.
\Fig{f:vprms_t_npart_comparison} shows that the computational time needed for
superparticle collisions is similar to or lower than the time needed for gas
hydrodynamics, particle dynamics, and drag forces, if the number of collision
partners is kept below approximately 100.

A side effect of reducing the number of collision partners is that the maximum
number of {\it collisions} is reduced accordingly. Therefore it be must
required that the boosted collision probability $P'=P (N-1)/N_{\rm max}$ is
always much smaller than unity. This must hold for all particle pairs. One can
use the maximum relative speed between any two particles within a grid cell to
estimate the smallest allowed $N_{\rm max}$ that keeps all $P'\ll1$.

Each swarm in our simulations represents $\hat{\rho}_{\rm p}/\rho_{\rm
g}\approx0.219$. The base probability for collision between two superparticle
swarms with random motion $\delta v/c_{\rm s}\sim0.025$ is $P=\delta
t/\tau_{\rm c} \sim 10^{-5}$ using \Eq{eq:tcoll_tfric_eq} and a typical
hydrodynamical time-step $\delta t$ at $64^3$ and $128^3$ resolution. The
maximum density reached in the simulations is $\rho_{\rm p}/\rho_{\rm
g}\approx3000$ (see \Fig{f:rhopmax_t_noshear}), giving $\approx13700$ particles
in the densest cells. We use $N_{\rm max}=100$ and thus the maximum boosted
probability is $P' \sim 10^{-3}$, safely in the regime where the collision
time-scale can be ignored when determining the numerical time-step of the code.

\section{From superparticles to inflated particles}
\label{s:inflated}

Consider a particle component of mass density $\rho_{\rm p}$. A superparticle
can maximally hold a particle number $\hat{N}$ (equivalently particle mass
density $\hat{\rho}_{\rm p}$) that covers the whole area of the grid cell,
\begin{equation}
  \hat{N} \sigma = \frac{\hat{\rho}_{\rm p}}{m_{\rm p}} (\delta x)^3 \sigma =
  \frac{\hat{\rho}_{\rm p}}{\rho_{\rm p}} \frac{(\delta x)^3}{\lambda} <
  (\delta x)^2 \, .
\end{equation}
Here $\sigma$ is the cross section of a swarm member and $\lambda$ is the mean
free path of physical particles in the system. This gives a maximum
superparticle mass density of
\begin{equation}
  \hat{\rho}_{\rm p} = \rho_{\rm p} \frac{\lambda}{\delta x} \, .
\end{equation}
At this mass density the Monte Carlo method breaks down because the
superparticle area is larger than a single grid cell (this is not taken into
account in the model because collisions are only considered when superparticles
share the same grid cell). The free path of a test particle encountering this
maximum density superparticle is
\begin{equation}
  \hat{\lambda} = \frac{1}{\hat{n} \sigma} = \frac{\rho_{\rm
  p}}{\hat{\rho}_{\rm p}} \lambda = \delta x \, ,
\end{equation}
using $\sigma = 1/(\hat{\lambda}\hat{n}) = 1/(\lambda n)$. Thus the maximum
area criterion coincides with the particle density where the free path is the
same as the grid spacing, giving a collision probability of unity when the
particle enters a grid cell occupied by a superparticle. This is in fact
equivalent to the inflated particle approach, i.e.\ that overlapping particles
always collide.

We still must show that the mean free path of the system is equal to the
physical mean free path. The total particle number in the box is
\begin{equation}
  N = \frac{\rho_{\rm p} L^3}{\hat{\rho}_{\rm p} (\delta x)^3} \, .
\end{equation}
This gives a mean free path for the ``grid point particles'' of
\begin{equation}
  \lambda' = \frac{L^3}{N \sigma} = \frac{\hat{\rho}_{\rm p} (\delta
  x)^3}{\rho_{\rm p} (\delta x^2)} = \frac{\hat{\rho}_{\rm p}}{\rho_{\rm p}}
  \delta x = \lambda \, .
\end{equation}
This shows how the superparticle Monte Carlo method smoothly transforms to the
inflated particle method when reducing the number of superparticles and
increasing their mass. At the point when the superparticle fills up its grid
cell, the collision probability approaches unity inside the cell and the mean
free path of the grid cell particles is equal to the mean free path of the
physical particles. At the same time one must only allow approaching particles
to collide, to avoid multiple collisions inside the grid cell. Of course, the
collision detection algorithm for these cubic particles is rather crude, but
the geometric effect of considering cubic rather than spherical particles is
minor.

\end{document}